\documentclass[]{spie}  

\usepackage{amsmath,amsfonts,amssymb}
\usepackage{wasysym}
\usepackage{graphicx}
\usepackage[colorlinks=true, allcolors=blue]{hyperref}
\usepackage[T1]{fontenc}
\usepackage{hyperref}

\title{Validating the Local Volume Mapper acquisition and guiding hardware}

\author[a]{Maximilian Häberle}
\author[a]{Thomas M. Herbst}
\author[a]{Peter Bizenberger}
\author[b]{Guillermo Blanc}
\author[a]{Florian Briegel}
\author[c]{Niv Drory}
\author[a]{Wolfgang Gässler}
\author[b]{Nick Konidaris}
\author[d]{Kathryn Kreckel}
\author[a]{Markus Kuhlberg}
\author[a]{Lars Mohr}
\author[e]{Eric Pellegrini}
\author[b]{Solange Ramirez}
\author[a]{Christopher Ritz}
\author[a]{Ralf-Rainer Rohloff}
\author[a]{Paula Stępień}

\affil[a]{Max-Planck-Institut für Astronomie, Königstuhl 17, 69117 Heidelberg, Germany}
\affil[b]{Observatories | The Carnegie Institution for Science, 813 Santa Barbara Street, Pasadena, CA 91101, USA}
\affil[c]{McDonald Observatory / Dept. of Astronomy, The University of Texas at Austin,2515 Speedway, Stop C1402 Austin, Texas 78712-1206, USA}
\affil[d]{Zentrum für Astronomie der Universität Heidelberg, Astronomisches Rechen-Institut, Mönchhofstr. 12-14, 69120 Heidelberg, Germany}
\affil[e]{Zentrum für Astronomie der Universität Heidelberg, Institut für Theoretische Astrophysik, Albert-Ueberle-Str. 2, 69120 Heidelberg, Germany}

\authorinfo{Further author information: (Send correspondence to Maximilian Häberle, E-mail: haeberle@mpia.de)}

\begin{document} 
\maketitle

\begin{abstract}
The Local Volume Mapper (LVM) project is one of three surveys that form the Sloan Digital Sky Survey V. It will map the interstellar gas emission in a large fraction of the southern sky using wide-field integral field spectroscopy. Four 16-cm telescopes in siderostat configuration feed the integral field units (IFUs).
A reliable acquisition and guiding (A\&G) strategy will help ensure that we meet our science goals. Each of the telescopes hosts commercial CMOS cameras used for A\&G.
In this work, we present our validation of the camera performance. Our tests show that the cameras have a readout noise of around 5.6~e- and a dark current of 21~e-/s, when operated at the ideal gain setting and at an ambient temperature of 20~°C. To ensure their performance at a high-altitude observing site, such as the Las Campanas Observatory, we studied the thermal behaviour of the cameras at different ambient pressures and with different passive cooling solutions.
Using the measured properties, we calculated the brightness limit for guiding exposures. With a 5 s exposure time, we reach a depth of $\sim$16.5 Gaia gmag with a signal-to-noise ratio (SNR)~$>$~5. Using Gaia Early Data Release 3, we verified that there are sufficient guide stars for each of the $\sim$25~000 survey pointings.
For accurate acquisition, we also need to know the focal plane geometry. We present an approach that combines on-chip astrometry and using a point source microscope to measure the relative positions of the IFU lenslets and the individual CMOS pixels to around 2~µm accuracy.

\end{abstract}

\keywords{Local Volume Mapper, CMOS sensors, Instrument Metrology, Point Source Microscope, Telescopes, Guiding, Sloan Digital Sky Survey V}

\section{INTRODUCTION}
The Sloan Digital Sky Survey V (SDSS-V) is an all-sky spectroscopic survey of <6 million objects, designed to decode the history of the Milky Way, reveal the inner workings of stars, investigate the origin of solar systems, and track the growth of supermassive black holes across the Universe\cite{2017arXiv171103234K}. The three individual programs within the SDSS-V are the Milky Way Mapper, the Black Hole Mapper, and the Local Volume Mapper (LVM)\cite{2020SPIE11447E..18K}.

The goal of the LVM project is to map the interstellar gas emission in a large fraction of the southern sky using wide-field integral field spectroscopy. During the  $\sim$4.5 years of the survey around 2500 deg$^2$ of sky will be mapped with a spectral resolution of around R$\sim$ 4000.

To achieve these goals, an entirely new facility is currently under construction at Las Campanas Observatory in Northern Chile. The LVM instrument consists of four 16-cm telescopes\cite{2020SPIE11445E..0JH} in siderostat configuration. Each of the telescopes is equipped with an integral field unit (IFU) that is connected to DESI like spectrographs\cite{2018SPIE10702E..7KP}, that have a spectral resolution of R$\sim$4000 and a wavelength range from 365~nm to 980~nm. One telescope conducts the scientific observations, while two telescopes will observe sky fields for sky-emission calibration, and the final telescope will observe multiple spectro-photometric calibration stars. Herbst el al. (paper. 12184-135, this conference) gives an up-to-date overview of the telescope construction and commissioning phase.

\section{Guiding and Acquisition cameras}
\subsection{Optical setup}
\begin{figure} [ht]
   \begin{center}
   \includegraphics[width=16cm]{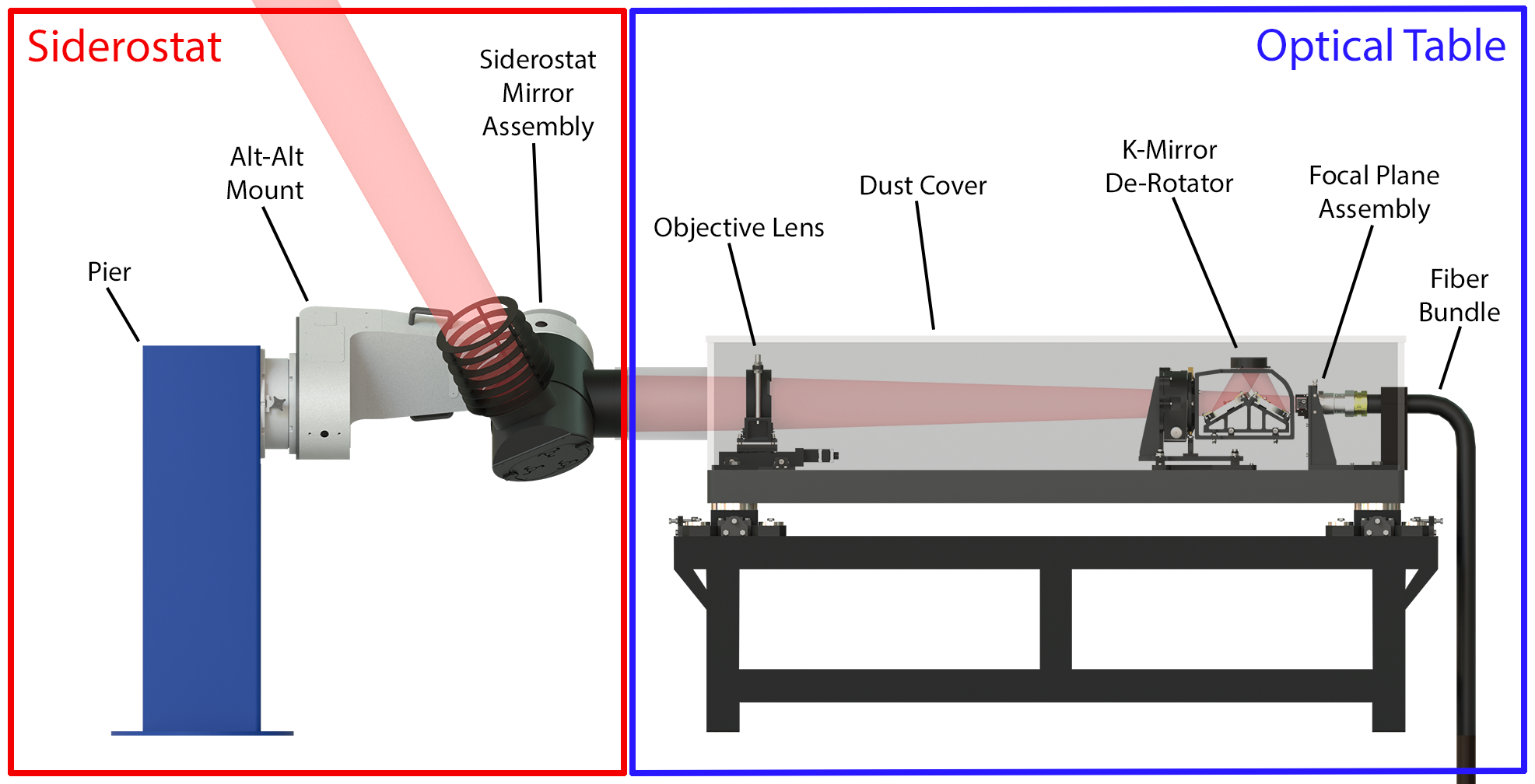}

   \end{center}
   \caption[table] 
   { \label{fig:table_setup} 
Schematic of the table mounted optical setup.}
   \end{figure}

The siderostats feed the light to a stable, table-mounted, optical setup (see Figure~\ref{fig:table_setup}). On the optical tables, the light is focused by a \diameter16.1~cm, f/11.42 triplet lens\footnote{Lanz et al. (paper 12184-218, this conference) provide a complete description of the objective lens design, manufacturing, and testing.}. After the lens, the beam is de-rotated by a K-mirror\footnote{See Kuhlberg et al. (paper 12184-261, this conference) for details on the K-mirror.},
before it reaches the IFU and the guiding system in the focal plane.
\subsection{Focal plane design}
There are three slightly different focal plane layouts for the different telescopes. The Science and the Sky telescopes are equipped with 2 A\&G-sensors and an IFU in the centre of the focal plane.  The science IFU has a hexagonal shape with 1801 microlenses, while the sky IFUs of the Sky telescopes are smaller with 59 and 60 microlenses. The microlenses have a diameter of 315~µm and a spacing of 330~µm.

\begin{figure} [ht]
   \begin{center}
   \includegraphics[height=3.7cm]{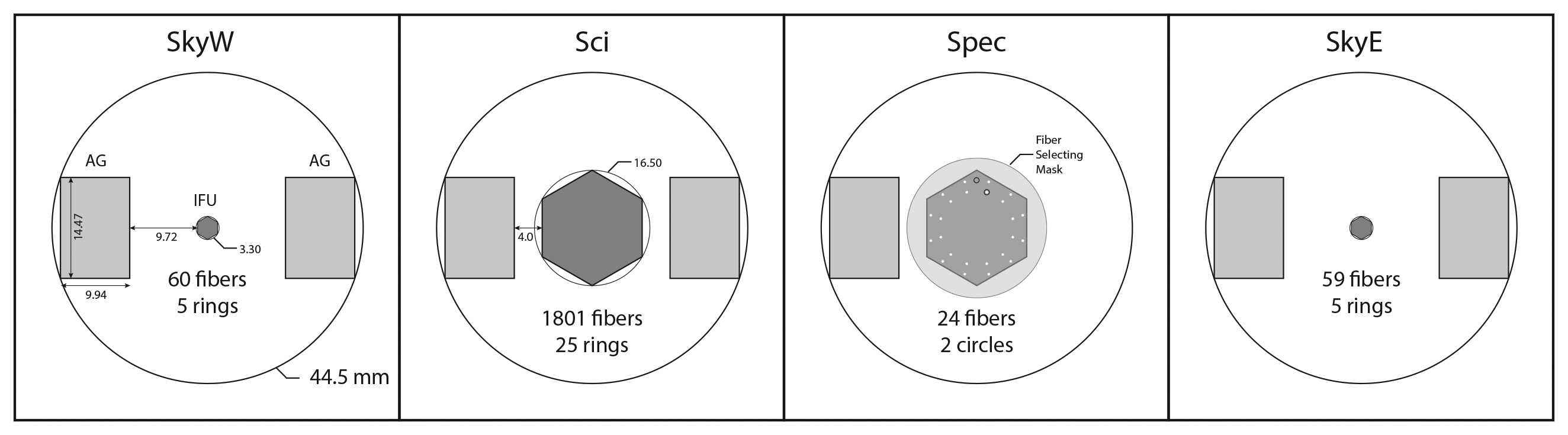}
   \includegraphics[height=3.6cm]{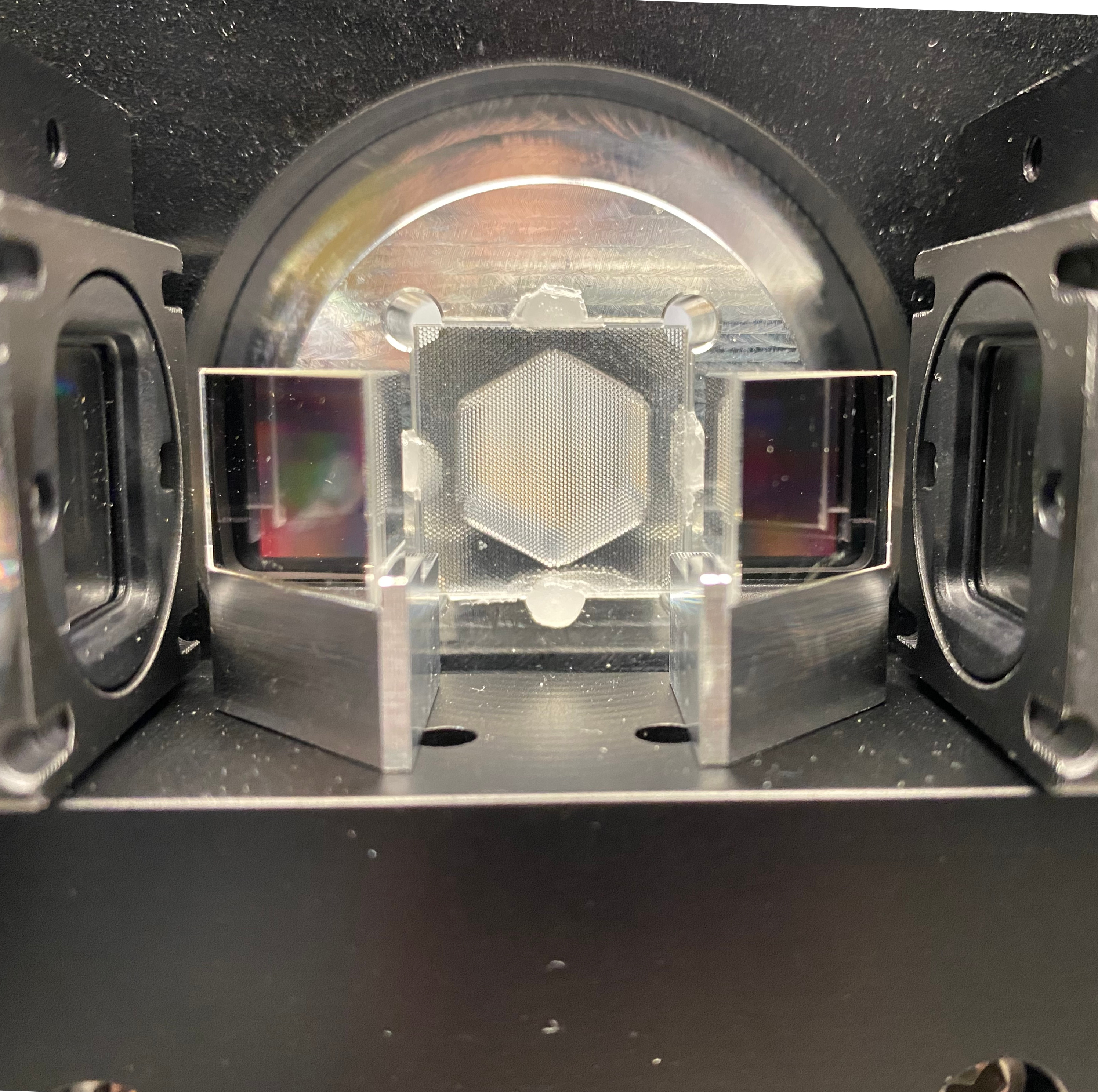}

   \end{center}
   \caption[table] 
   { \label{fig:fp_layout} 
On the left, we show a schematic of the different focal plane layouts of the 4 LVM telescopes. Note that for the A\&G sensors the positions shown are after the light is reflected by the 45$^\circ$ mirrors. The photo on the right shows a test setup for the focal plane metrology. The full Science IFU is substituted by a dummy microlens array.}
   \end{figure}

The Spectrophotometric telescope is equipped with a single A\&G sensor and hosts a rotation mask shutter mechanism, that allows illumination of individual fibres. Instead of a fully populated IFU, only 24 fibres are mounted.  A schematic of the different layouts appears in Figure~\ref{fig:fp_layout}.

In all cases, the sensors of the A\&G cameras are mounted orthogonal to the plane of the fibers, and a 45$^\circ$~mirror reflects the light to the sensors. A test setup that is very similar to the final design appears in Figure~\ref{fig:fp_layout}.
\subsection{CHARACTERISATION OF THE NOISE PROPERTIES AND THE THERMAL PERFORMANCE OF THE ACQUISITION AND GUIDING CAMERAS}
\subsubsection{The Cameras}
In the past, active pixel complementary metal-oxide semiconductor (CMOS) sensors have not been a viable alternative to charge-coupled device (CCD) sensors for astronomical applications, due to higher read-out noise and low quantum efficiency. However, in recent years, new developments and technology improvements have changed this\cite{2006SPIE.6276E..0MJ,2013SPIE.8659E..02J,2017JInst..12C7008J} and there are various CMOS high performance devices commercially available.

We use commercial \text{FLIR Blackfly S BFS-PGE-16S7M} cameras as A\&G cameras. These compact cameras have a GigE interface and use power over Ethernet, which means only a single cable per camera is needed. The cameras are equipped with the \textit{Sony IMX432} CMOS sensor\footnote{Datasheet: \url{https://www.sony-semicon.co.jp/products/common/pdf/IMX432LLJ_LQJ_Flyer.pdf}}. This monochrome sensor has an effective pixel count of 1608x1104 and a physical size of 14.4~mm~x~9.9~mm. The pixel pitch is 9~µm. As these CMOS devices are typically used in non-astronomical applications (such as high-frame rate machine vision), the relevant quantities for photometric applications (read noise, dark current, gain) are not stated on the datasheet. We therefore characterised them in the laboratory.
\subsubsection{Gain and Read noise measurements}
The photons hitting the detector generate electrons, but the cameras deliver values in analog-digital-converter units (ADUs). The gain is the (adjustable) conversion factor between the number of measured electrons and the ADUs.

\begin{figure} [ht]
   \begin{center}
   \includegraphics[width=16cm]{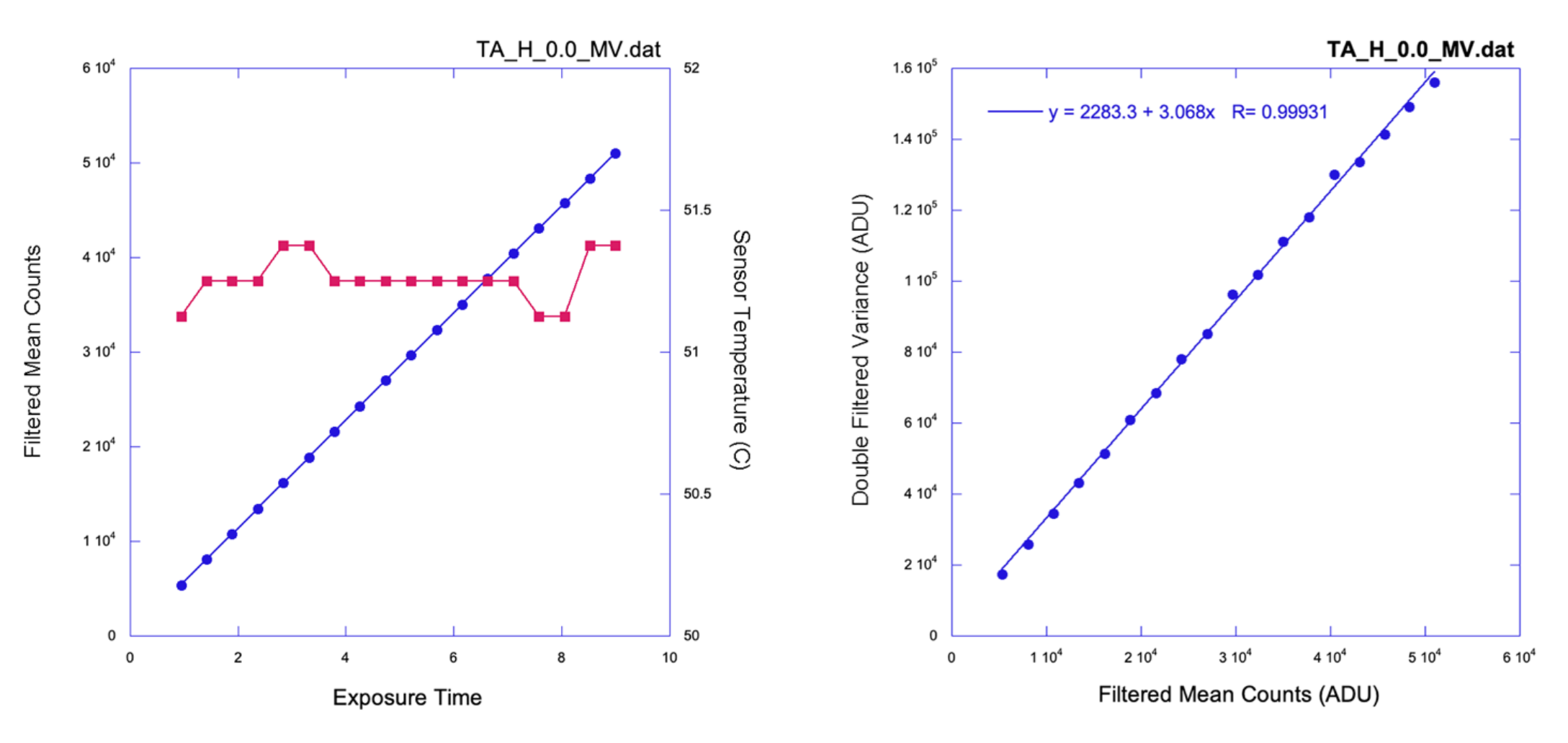}

   \end{center}
   \caption[table] 
   { \label{fig:mean_variance} 
The left plot shows how the mean signal increases linearly with the exposure time. As the sensor temperature also affects the performance, we logged it during the measurement. The right plot shows the variance plotted against the mean counts.}
   \end{figure}

The classical method to determine the gain value is to do a so called mean-variance test: As photons (and the generated electrons) obey Poisson statistics, the standard deviation of measured electrons is related to their count by $\sigma_e = \sqrt{N_e}$. This leads to a linear relation between the variance in ADU ($\sigma^2_{ADU}$) and the signal:
\begin{equation}
\sigma^2_e = (Gain\cdot\sigma_{ADU})^2=N_e=Gain\cdot N_{ADU}    
\end{equation}
The slope of the linear relation between mean (ADU-)counts per pixel and the variance gives the gain. To make the measurment, we linearly increased exposure time with constant illumination until the sensors reached saturation. For each exposure time/signal level, we saved 20 images, allowing us to determine the variance per pixel (which unlike the total of the image is not affected by pixel to pixel variations). After that, we performed a linear fit function to the mean-variance points. An example for such a measurement is shown in Figure~\ref{fig:mean_variance}.

At the same time, the zero-signal variance (or Y-intercept of the linear fit to the mean-variance relation) gives a measure of the read noise.
\subsubsection{Results and determination of the optimal gain}
\label{subsec:noise}
The cameras offer various gain settings, separated into \textit{High Conversion Gain (HCG)} and \textit{Low Conversion Gain (LCG)} modes and with an adjustable gain value between 0 and 48. We performed the mean-variance test for seven gain values in the two modes. The results for the different settings appear in Table~\ref{tab:gain_results}. The gain setting LCG~48 produced questionable results and is excluded from the table. In general, the \textit{HCG} modes show significantly lower read noise, and increasing the gain of the camera leads to lower read noise. For our LVM use case, a gain setting of 5 in the \textit{HCG} mode is a good compromise between dynamic range (with a gain of 0.177e$^-$/ADU we reach saturation at $\sim$12~000e$^-$) and read noise ($\sigma_R=5.6~e^-$), and we use this as the baseline for our guide star brightness calculations (\ref{subsec:sn}).
   \begin{table}[ht]
\caption{Results for gain and read noise for different gain settings of the cameras. All measurements were made with a sensor temperature of $\sim$51$^\circ$~C.}
\label{tab:gain_results}
\begin{center} 
\begin{tabular}{llllllll}
\hline
\multicolumn{1}{|l}{}                    &       &       & \multicolumn{5}{l|}{High conversion gain:}                         \\ \hline
\multicolumn{1}{|l}{Gain Setting}        & 0     & 5     & 15     & 25     & 35      & 45      & \multicolumn{1}{l|}{48}      \\ \hline
\multicolumn{1}{|l}{Gain {[}e-/ADU{]}}   & 0.326 & 0.177 & 0.0562 & 0.0180 & 0.00575 & 0.00178 & \multicolumn{1}{l|}{0.00125} \\
\multicolumn{1}{|l}{Read Noise {[}e-{]}} & 15.6  & 5.62  & 3.79   & 2.29   & 2.98    & 2.67    & \multicolumn{1}{l|}{2.59}    \\ \hline
                                         &       &       &        &        &         &         &                              \\ \hline
\multicolumn{1}{|l}{}                    &       &       & \multicolumn{5}{l|}{Low conversion gain:}                          \\ \hline
\multicolumn{1}{|l}{Gain Setting}        & 0     & 5     & 15     & 25     & 35      & 45      & \multicolumn{1}{l|}{48}      \\ \hline
\multicolumn{1}{|l}{Gain {[}e-/ADU{]}}   & 1.59  & 0.878 & 0.289  & 0.0918 & 0.0294  & 0.00889 & \multicolumn{1}{l|}{-}       \\
\multicolumn{1}{|l}{Read Noise {[}e-{]}} & 24.5  & 12.0  & 13.6   & 12.9   & 12.9    & 11.8    & \multicolumn{1}{l|}{-}       \\ \hline
\end{tabular}
\end{center}
\end{table}
\subsubsection{Dark current measurements}
\begin{figure} [ht]
   \begin{center}
   \includegraphics[height=7.5cm]{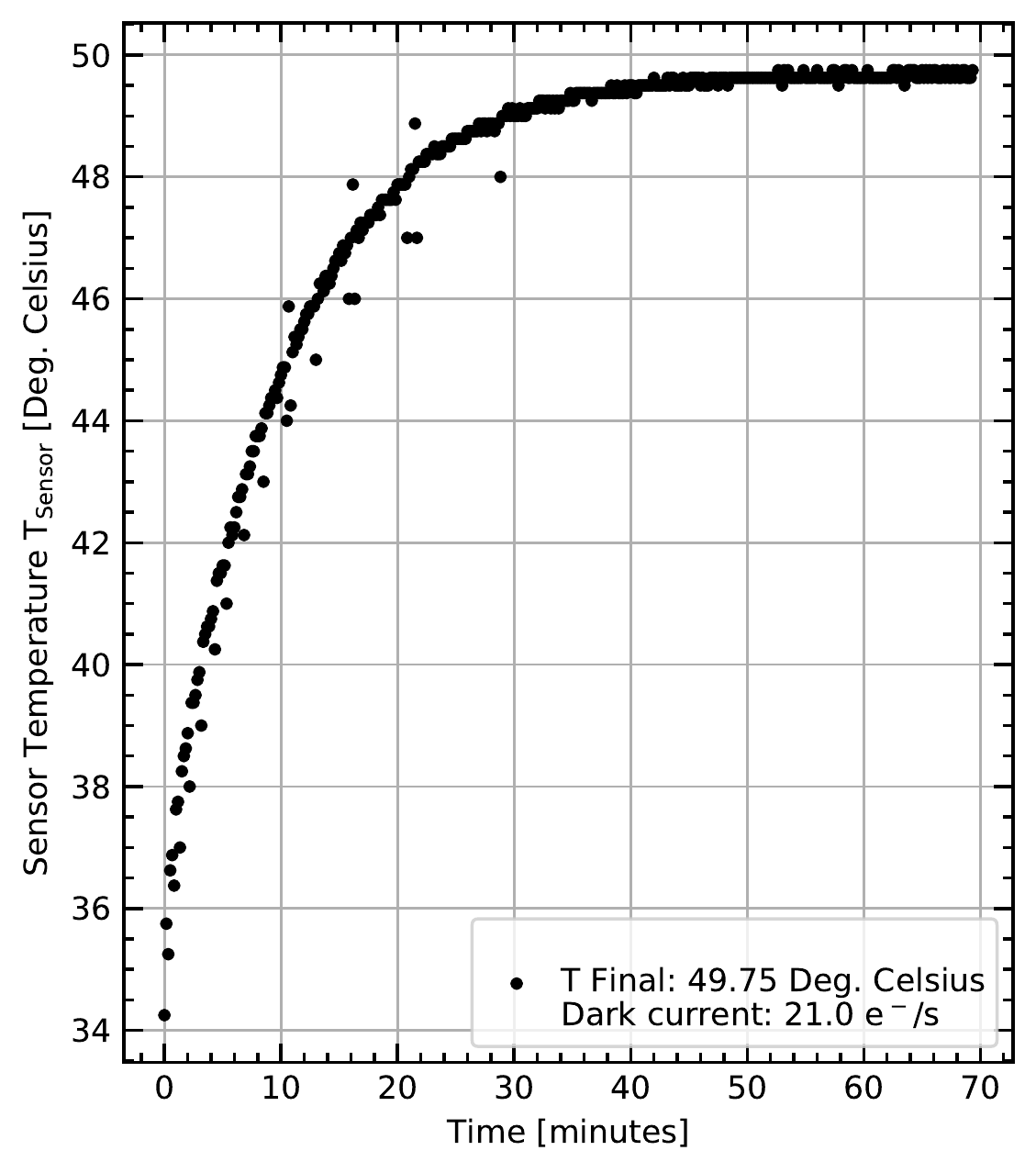}
   \includegraphics[height=7.5cm]{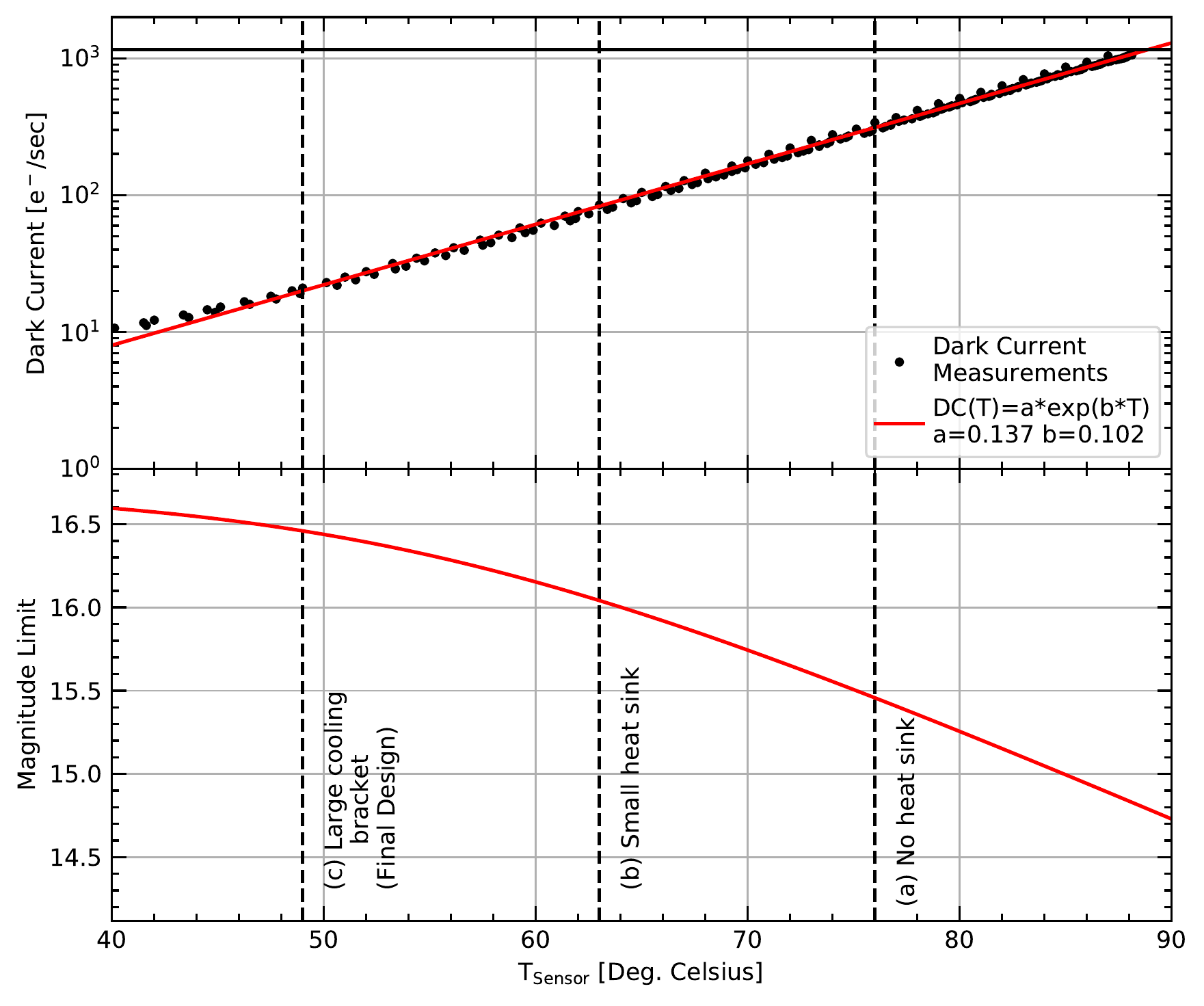}

   \end{center}
   \caption[position angles] 
   { \label{fig:dark_current} 
The left panel shows how one of the A\&G cameras heats up under laboratory conditions and reaches its equilibrium temperature after around 45 minutes. The right panel shows the exponential dependence of the dark current on the temperature (top) and its effect on the magnitude limit (bottom, see also section \ref{subsec:sn}).}
   \end{figure}

The dark current was measured by taking so-called dark-frames. We kept the sensor in complete darkness by dimming the lights in the laboratory and sealing the sensor with a plastic cap. The dark frames were determined while the cameras were warming up, using a 10~s exposure time. In this way, we also measure the temperature dependence of the dark current (see next section and Figure~\ref{fig:dark_current}). We determined the median image value and used the measured gain and the exposure time to convert the image counts to electrons/second.
\subsubsection{Temperature behaviour and optimisation of passive cooling}
\begin{figure} [ht]
   \begin{center}
   \includegraphics[width=12cm]{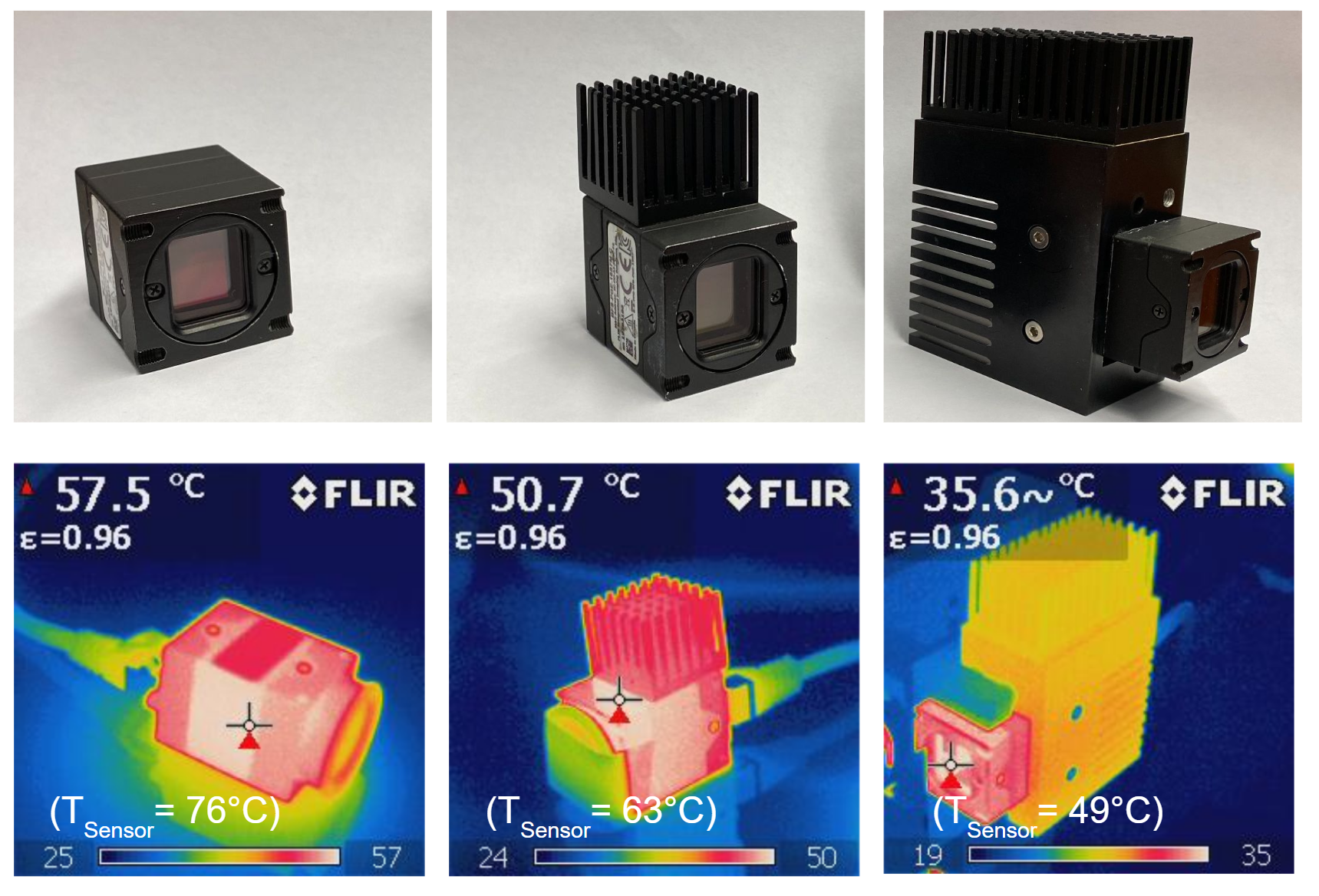}

   \end{center}
   \caption[heatsinks] 
   { \label{fig:heatsinks} 
The upper row shows photos of the different heatsink configurations which we tested. The lower row shows corresponding thermal images, which we used to verify, whether the heatsinks are working efficiently.}
   \end{figure}

The dark current of a silicon detector shows a strong dependence on temperature. Even though the average power consumption of each of the cameras is just 3~W (including the Ethernet interface), the sensors themselves reach a much higher temperature than their surroundings, due to their small physical size and therefore heat dissipation. Our camera model has an onboard-temperature sensor close to the imaging sensor. In addition to this, we also measured the surface temperature of the camera housing using a thermal infrared camera (see Figure~\ref{fig:heatsinks}).

Under laboratory conditions (ambient temperature around 20°~C), a bare camera reached a sensor temperature of around 75$^\circ~C $, leading to a dark current of around 300~$e^-/s$. These conditions would affect the sensitivity of the guide images in a very negative way (see Figure~\ref{fig:dark_current}. The details of the signal-to-noise ratio (SNR) calculation are explained in \ref{subsec:sn}). For this reason, many cameras used for astronomical applications are actively cooled (either using thermoelectric cooling with Peltier devices, or if even lower temperatures are required, cryogenics such as liquid nitrogen). Due to the limited space in the focal plane, and to avoid additional complexity, we aimed to reduce the temperature of the cameras to an acceptable level using only passive cooling. We started with a small commercial (25~mm)$^2$ heatsink which led to a decrease of the sensor temperature of 12$^\circ$~C (to around 63$^\circ$~C). To further reduce the temperature we experimented with custom cooling brackets with additional heatsinks and fins. Beside the heat dissipation, the cooling brackets also serve as the adaptor to mount the cameras to the Focal Plane Assembly. An overview off the different tested setups appears in Figure~\ref{fig:heatsinks}. With the final cooling brackets, the equilibrium temperature of the sensors is $\sim$49$^\circ$~C, giving a dark current of 21~$e^-/s$ (a factor of $\sim$15 lower than the initial setup). This is of the same order as the sky-brightness in a full-moon-night; further reducing the dark current would therefore only have a small influence on the SNR of the stars (see \ref{subsec:sn}).
\subsubsection{Influence of ambient pressure}
\begin{figure} [ht]
   \begin{center}
   \includegraphics[width=1.0\textwidth]{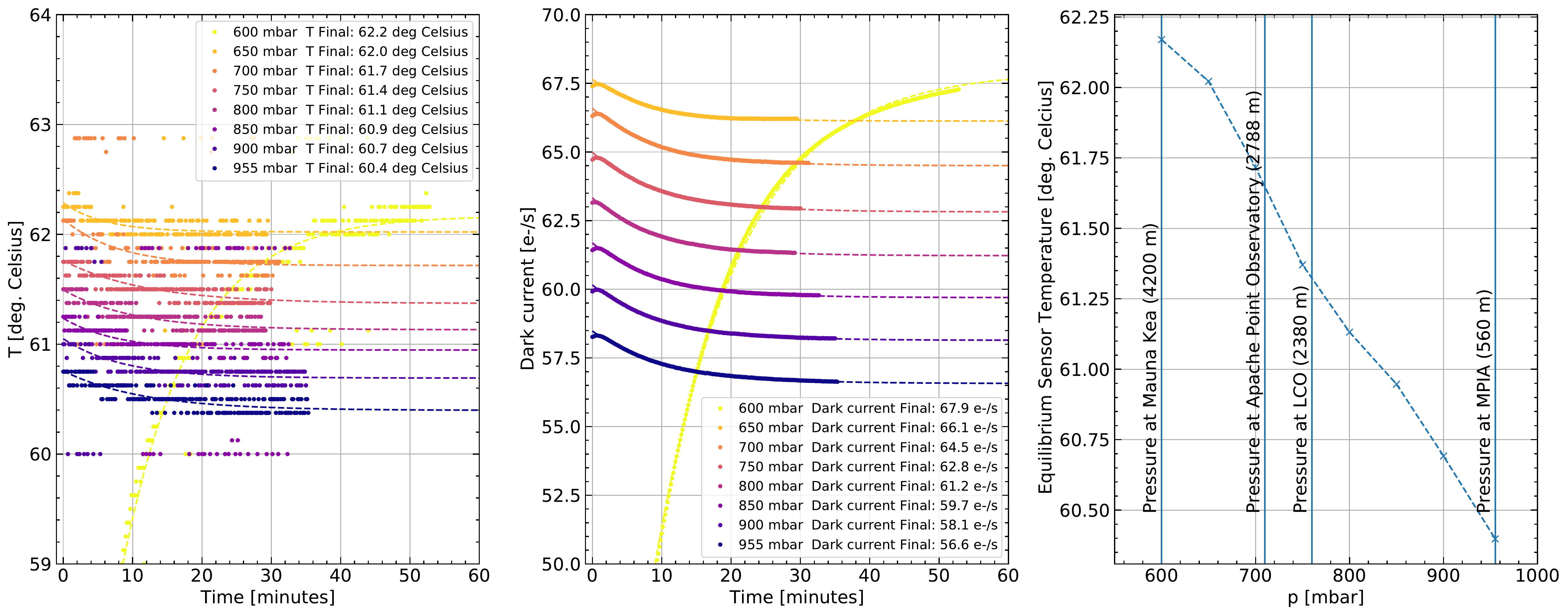}
   \end{center}
   \caption[pressure] 
   { \label{fig:pressure} 
The left plot shows a time series of how the camera sensor temperature reaches its equilibrium for different simulated ambient pressures. The middle plot shows the corresponding dark current. The plot on the right shows the equilibrium temperature plotted against the pressure. The vertical lines mark the pressure at a few representative observing locations.}
   \end{figure}

As air convection plays is an important mechanism for heat dissipation, we expect a worse cooling performance for the A\&G cameras at high-altitude observing sites where the ambient pressure is lower than in our close-to-sea-level laboratory. To ensure a sufficient cooling performance also at the observatory, we simulated a lower ambient pressure by placing the camera inside a vacuum vessel. We lowered the pressure to a minimum of around 600~mbar, similar to what would be expected at Mauna Kea at 4200~m elevation, one of the highest astronomical observing sites. Las Campanas observatory is situated at 2380~m elevation, with a typical pressure of 760~mbar. We waited until the sensor temperature of the camera reached its equilibrium and studied how the final temperature changes with pressure. The results are shown in Figure~\ref{fig:pressure}. As predicted, we see higher temperatures at lower pressures, but the difference is small: between our laboratory and LCO, we only expect a difference of around 1$^\circ$~C. Please note that these tests were performed with only a small heat-sink and not the final large cooling-bracket, but we expect a similar trend for the final cooling solution.

\section{VERIFYING THE NUMBER OF GUIDE STARS FOR EACH SURVEY POINTING}
\subsection{Introduction}
The LVM Survey aims to map the entire Galaxy - from the crowded inner regions to the relatively empty zones at higher Galactic latitudes. The stellar density in these different fields varies by orders of magnitude, and we have to ensure a sufficient number of guide stars for each of the $\sim$25~000 planned survey pointings.

Note that the symmetry of the focal plane (Figure~\ref{fig:fp_layout}) allows us to rotate the field of view by +60$^\circ$ or $-$60$^\circ$, in case a pointing would not have a sufficient number of guide stars. Figure~\ref{fig:sky_map} demonstrates the three different possible field rotation angles for an example pointing. 
\subsection{Throughput calculations}
\label{subsec:throughput}
\begin{figure} [ht]
   \begin{center}
   \includegraphics[width=0.8\textwidth]{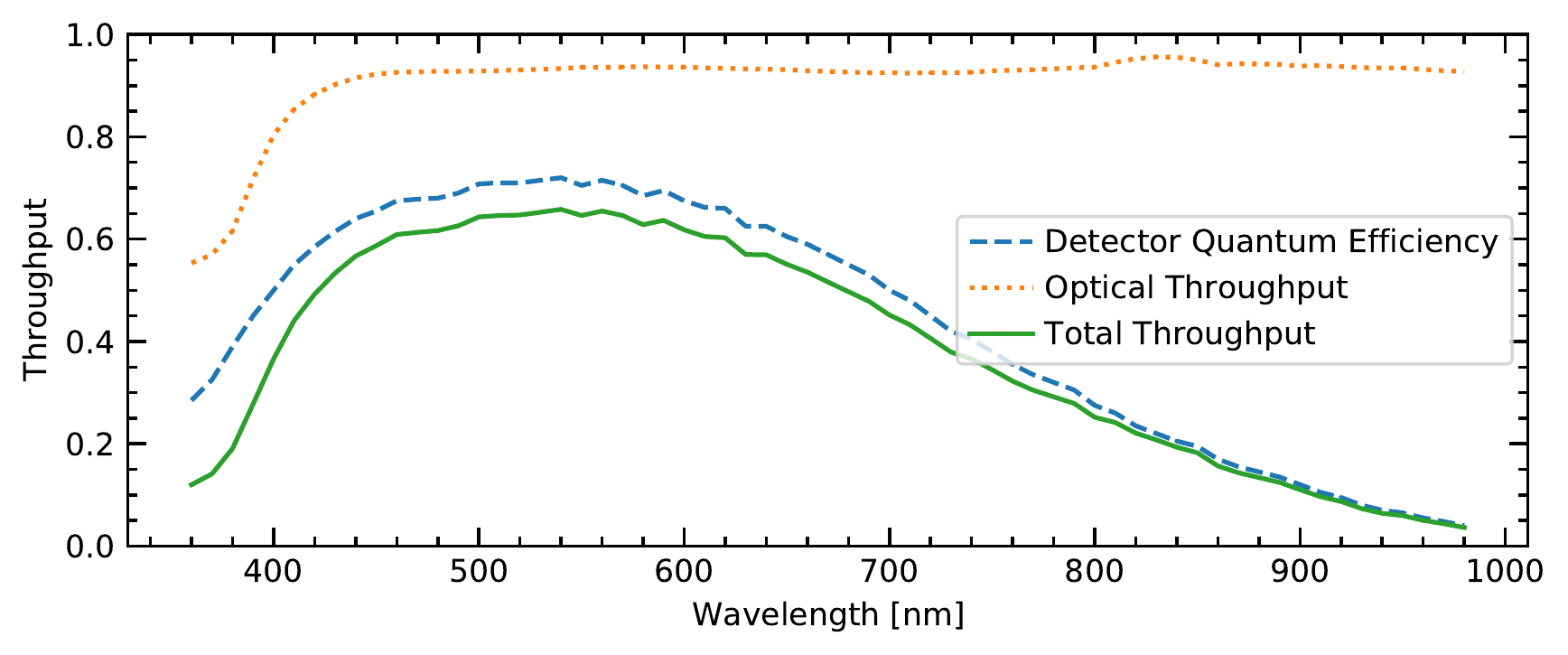}

   \end{center}
   \caption[throughput] 
   { \label{fig:throughput} 
The wavelength dependence of the LVM guiding system throughput for the science and sky telescopes.}
   \end{figure}

To determine the systems sensitivity, we have to take the full system throughput into account. To do so, we multiply the reflectivity of all reflective surfaces by the throughput of the lens and the quantum efficiency of the detector. The resulting wavelength dependent throughput function appears in Figure~\ref{fig:throughput}. We used the pyphot-package\footnote{\url{https://mfouesneau.github.io/pyphot/index.html}} to calculate the flux of Vega given our throughput. The resulting flux of Vega is $2.17\cdot10^6~e^-s^{-1}cm^{-2}$. With a telescope area of $\sim206$~cm$^2$, this gives us a total electron rate of $4.47\cdot10^8~e^-s^{-1}$ and a zeropoint of -21.6.

\subsection{Estimating the signal-to-noise ratio for guide stars}
\label{subsec:sn}
\begin{figure} [ht]
   \begin{center}
   \includegraphics[width=16cm]{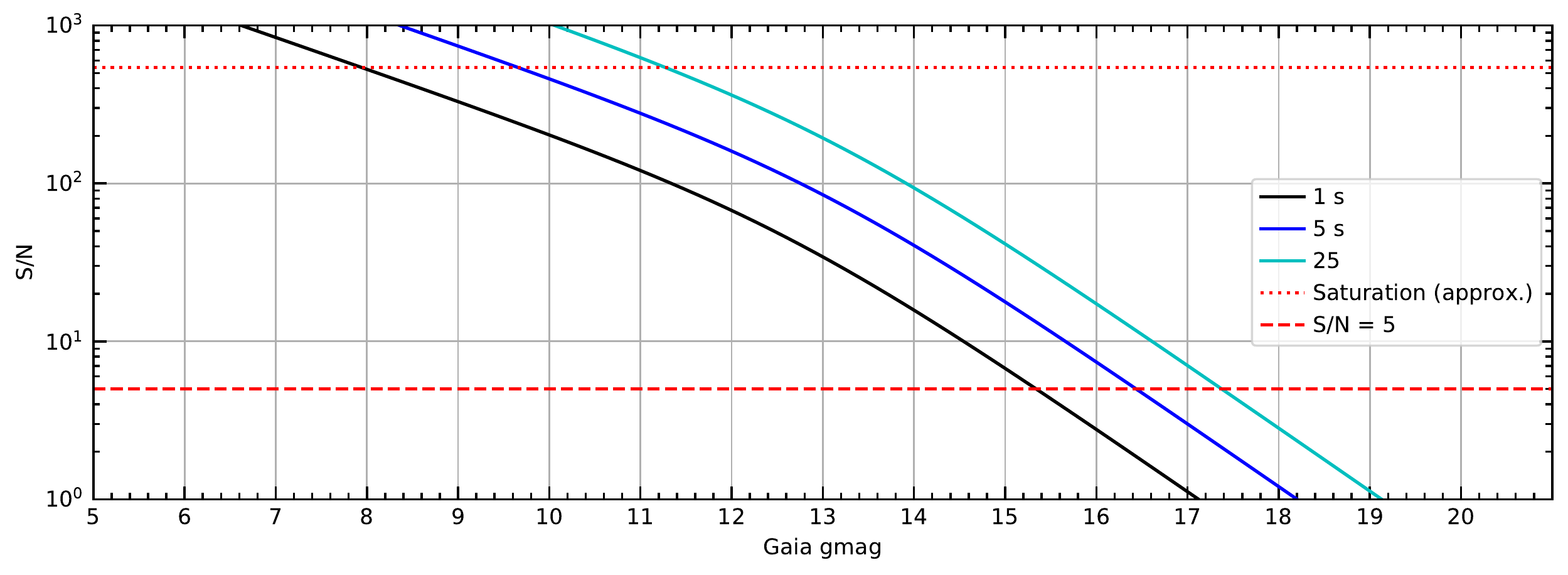}

   \end{center}
   \caption[synthetic image] 
   { \label{fig:sn_plot} 
SNR estimates for different guide star magnitudes and exposure times.
Those plots assume the as-measured read-out noise and the dark current with the passively cooled camera and a full moon night sky brightness}
   \end{figure}

To calculate the signal-to-noise ratio (SNR) for the guide stars, we assume a $N_{Pixel}=(7~\rm pixel)^2 = 49$ pixel guide window. This should capture all of the light, as the combined (seeing+diffraction+optical) requirement on full-width half maximum (FWHM) of the point-spread-function (PSF) size is 3.5~arcsec which is around 3.5 pixel.\\
The individual components entering the SNR calculation are:
\begin{description}
    \item[Signal level] We calculate the flux of electrons expected from a source with a specific magnitude using the zero point calculated in \ref{subsec:throughput}. The resulting number of electrons then is: $N_e = t_{exp}\cdot f_{\star}$.
    \item[Shot noise of source] The number of electrons is Poisson distributed with the standard deviation of the number electrons being: $\sigma_e=\sqrt{N_e}$.
    \item[Dark current] As measured in the laboratory, we use a dark current of $f_{Dark}=21~e^-/s/pixel$. This occurs for an A\&G sensor temperature of 49$^\circ$~C.
    \item[Sky Background] We calculate the sky background using the LVM exposure time calculator\footnote{\url{https://github.com/sdss/lvmetc_lco_script}}. In a full moon night (i.e. the worst case), we have a sky background of $f_{Sky}=28~e^-/s/pixel$.
    \item[Readout noise] As measured in the laboratory for our preferred gain setting (see \ref{subsec:noise}), we use a read noise of $\sigma_R=5.6e^-$.
\end{description}
We calculate the total noise by quadratically adding the different contributions. This leads us to the following expression:
\begin{equation}
SNR=\frac{t_{exp}\cdot f_{\star}}{\sqrt{N_{Pixel}\cdot(\sigma_R^2+t_{exp}\cdot(\frac{f_{\star}}{N_{Pixel}}+f_{Dark}+f_{Sky}))}}    
\end{equation}
Figure~\ref{fig:sn_plot} shows how the calculated SNR depends on the guide stars magnitude and on the exposure time. With the baseline exposure time of 5~s, we reach a SNR of at least 5 for stars with gmag$<$16.5.

\subsection{Querying the number of guide stars using \textit{Gaia} EDR3}   
We create a list of the planned survey pointings using the survey simulation tools\footnote{\url{https://github.com/sdss/lvmsurveysim}}.
In the relevant magnitude range for our small telescopes, the \textit{Gaia} Early Data Release 3 (EDR3)\cite{2021A&A...649A...1G} offers an essentially complete\cite{2021A&A...649A...5F} all-sky catalogue of stars. To determine a list of potential guide stars we performed the following steps:
\begin{enumerate}
    \item Determine the pointing coordinates from the Survey Planning Tools
    \item Query 0.69 deg around pointing coordinates using a local copy of the \textit{Gaia} catalogue. This query (which in principle has to choose from almost two billion Gaia stars) is made efficient by splitting the \textit{Gaia} catalogue into 50 000 HEALPix tiles\cite{Zonca2019, 2005ApJ...622..759G}\footnote{\url{http://healpix.sourceforge.net}} (Order 6).
    \item Convert angular coordinates into metric focal plane coordinates using an image scale of 8.92~µm/arcsec.
    \item Test which stars fall into the rectangular region of the two A\&G sensors for the three possible position angles (See Figure~\ref{fig:sky_map}).
\end{enumerate}
After the potential guide stars have been identified, we simply count the number of stars per sensor for different magnitude cutoffs. 
\subsection{Results}
\begin{figure} [ht]
   \begin{center}

   \includegraphics[width=0.8\textwidth]{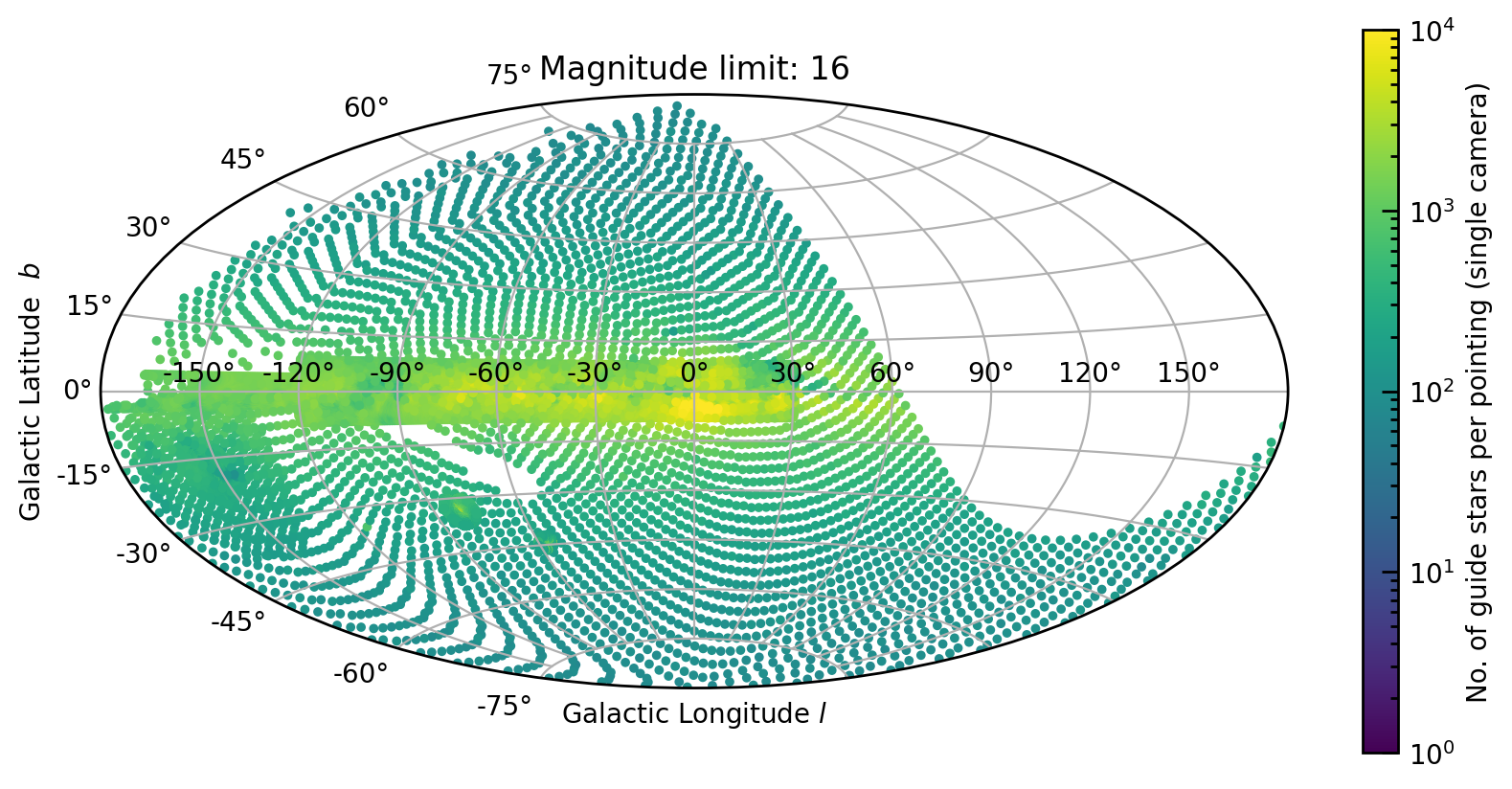}
   \includegraphics[width=0.4\textwidth]{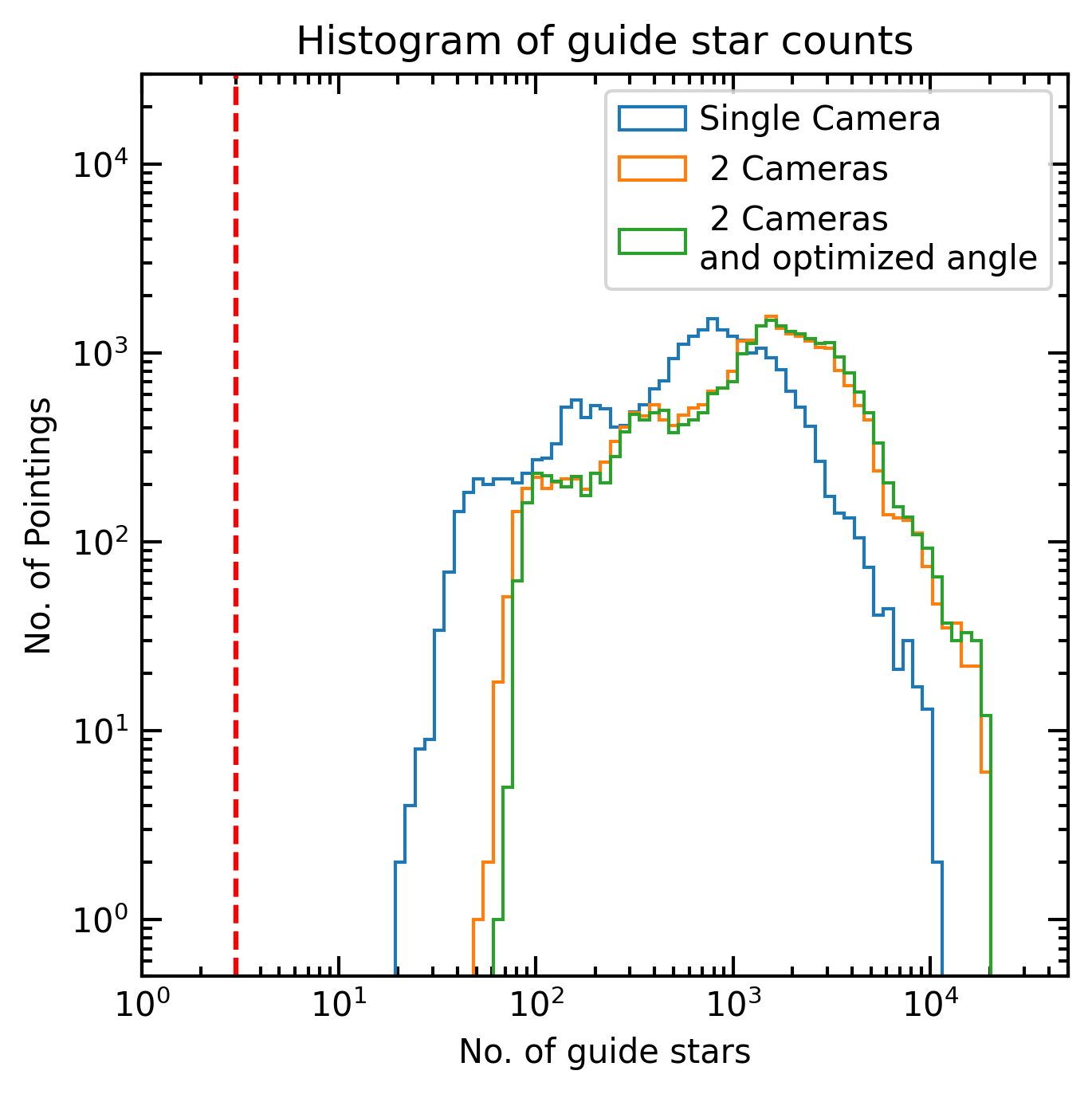}
   \includegraphics[width=0.4\textwidth]{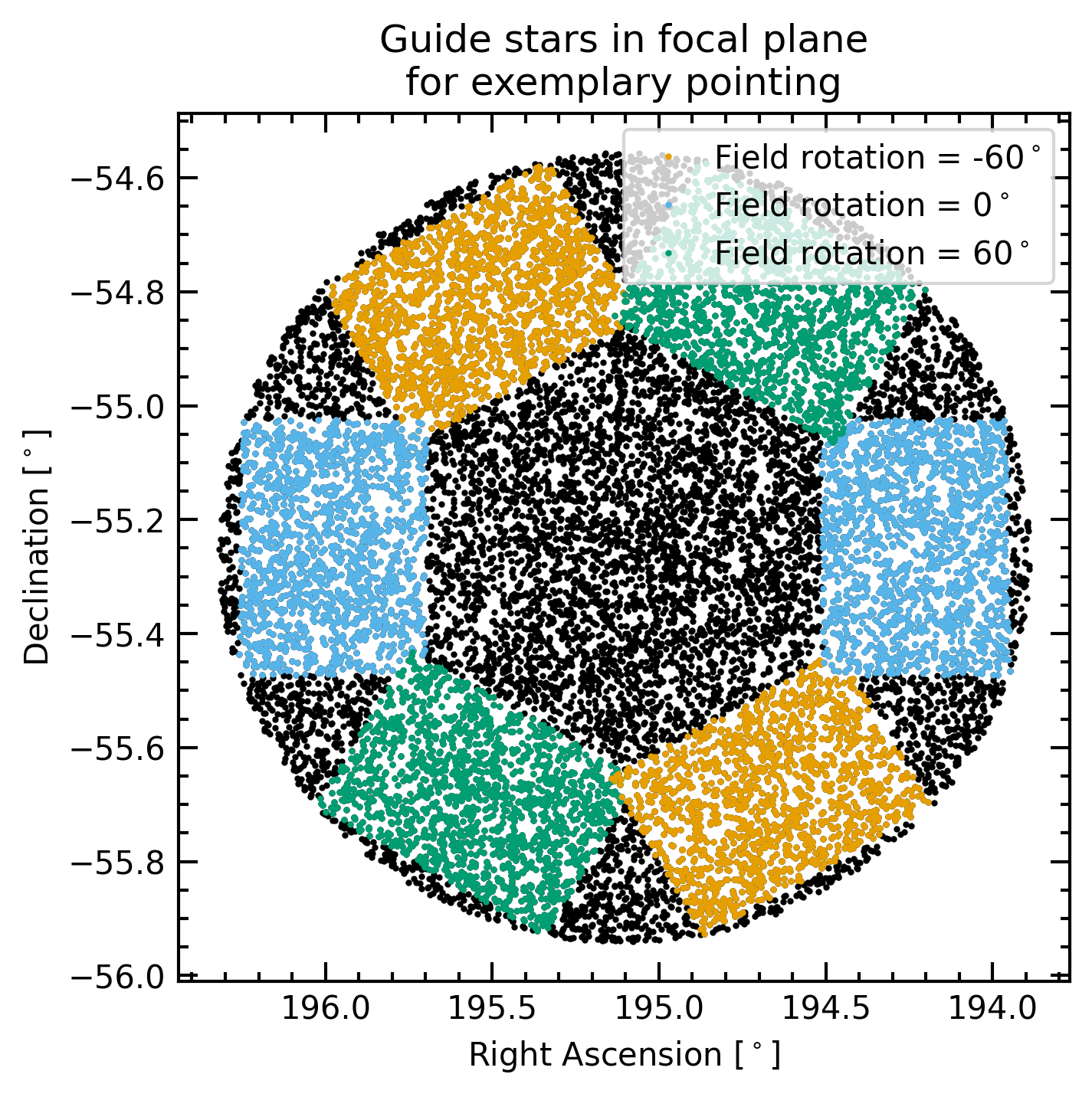}
   \end{center}
   \caption[sky map] 
   {\label{fig:sky_map}Top: A full-sky map in Galactic coordinates. The dots mark the different pointings of the survey and are color-coded with the number of available guide stars brighter than gmag=16. The imprint of the Milky Way with its dust clouds is clearly visible. Lower left: Histogram of the number of available guide stars brighter than gmag=16 for the different A\&G camera configurations. The dashed vertical line marks the critical number of 3 guide stars per pointing. All pointings have more stars than that. Lower right: The guide stars within the field of view for an example pointing at a moderately crowded field. The black dots indicate stars brighter than gmag < 16. The coloured dots mark stars within the field of view of the two A\&G Cameras at the 3 possible field rotation angles.}
   \end{figure}

The minimum, maximum and median number of guide-stars for various magnitude limits appears in Table~\ref{tab:guide_stars}. Figure~\ref{fig:sky_map} shows a full-sky map, color-coded with the number of available guide stars brighter and a histogram of the number of available guide stars for the representative magnitude limit of gmag$<$16.

Based on the SNR calculations (Figure~\ref{fig:sn_plot}) we can assume a magnitude limit of 16.5 mag in the brightest (full moon) nights. Even when adding more than a full magnitude as reserve, the minimum number of stars brighter than gmag~$=$~15 on a single camera chip is 6. Therefore, we can conclude that our guiding strategy will work for all planned survey pointings, and that no further mitigation strategies (such as longer exposure times, active camera cooling, optimisation of pointing coordinates, etc.) are necessary. 

\begin{table}[ht]
\caption{Results of the guide star number analysis. For a magnitude limit of $gmag < 15$, there is a sufficient number of guide stars even for a single camera.} 
\label{tab:guide_stars}
\begin{center}       
\begin{tabular}{l|lll|lll|lll|}
\cline{2-10}
                                 & \multicolumn{3}{l|}{Single camera}         & \multicolumn{3}{l|}{\begin{tabular}[c]{@{}l@{}}Two cameras in\\ standard orientation\end{tabular}} & \multicolumn{3}{l|}{\begin{tabular}[c]{@{}l@{}}Two cameras with\\ optimal orientation\end{tabular}} \\ \hline
\multicolumn{1}{|l|}{}           & \multicolumn{3}{l|}{\# No. of guide stars} & \multicolumn{3}{l|}{\# No. of guide stars}                                                         & \multicolumn{3}{l|}{\# No. of guide stars}                                                          \\
\multicolumn{1}{|l|}{Mag. Limit} & Min.         & Med.         & Max.         & Min.                           & Med.                           & Max.                             & Min.                            & Med.                           & Max.                             \\ \hline
\multicolumn{1}{|l|}{14.0}       & 2            & 140          & 1081         & 11                             & 288                            & 1966                             & 20                              & 313                            & 1967                             \\
\multicolumn{1}{|l|}{15.0}       & 6            & 322          & 3908         & 27                             & 660                            & 6683                             & 39                              & 709                            & 7120                             \\
\multicolumn{1}{|l|}{16.0}       & 20           & 704          & 10747        & 53                             & 1456                           & 18464                            & 66                              & 1554                           & 19797                            \\
\multicolumn{1}{|l|}{17.0}       & 41           & 1491         & 33994        & 98                             & 3069                           & 58460                            & 111                             & 3272                           & 61546                            \\
\multicolumn{1}{|l|}{18.0}       & 63           & 2991         & 66610        & 155                            & 6235                           & 125183                           & 175                             & 6671                           & 125183                           \\ \hline
\end{tabular}
\end{center}
\end{table}

\subsection{Synthetic images}
\begin{figure} [ht]
   \begin{center}
   \includegraphics[width=16cm]{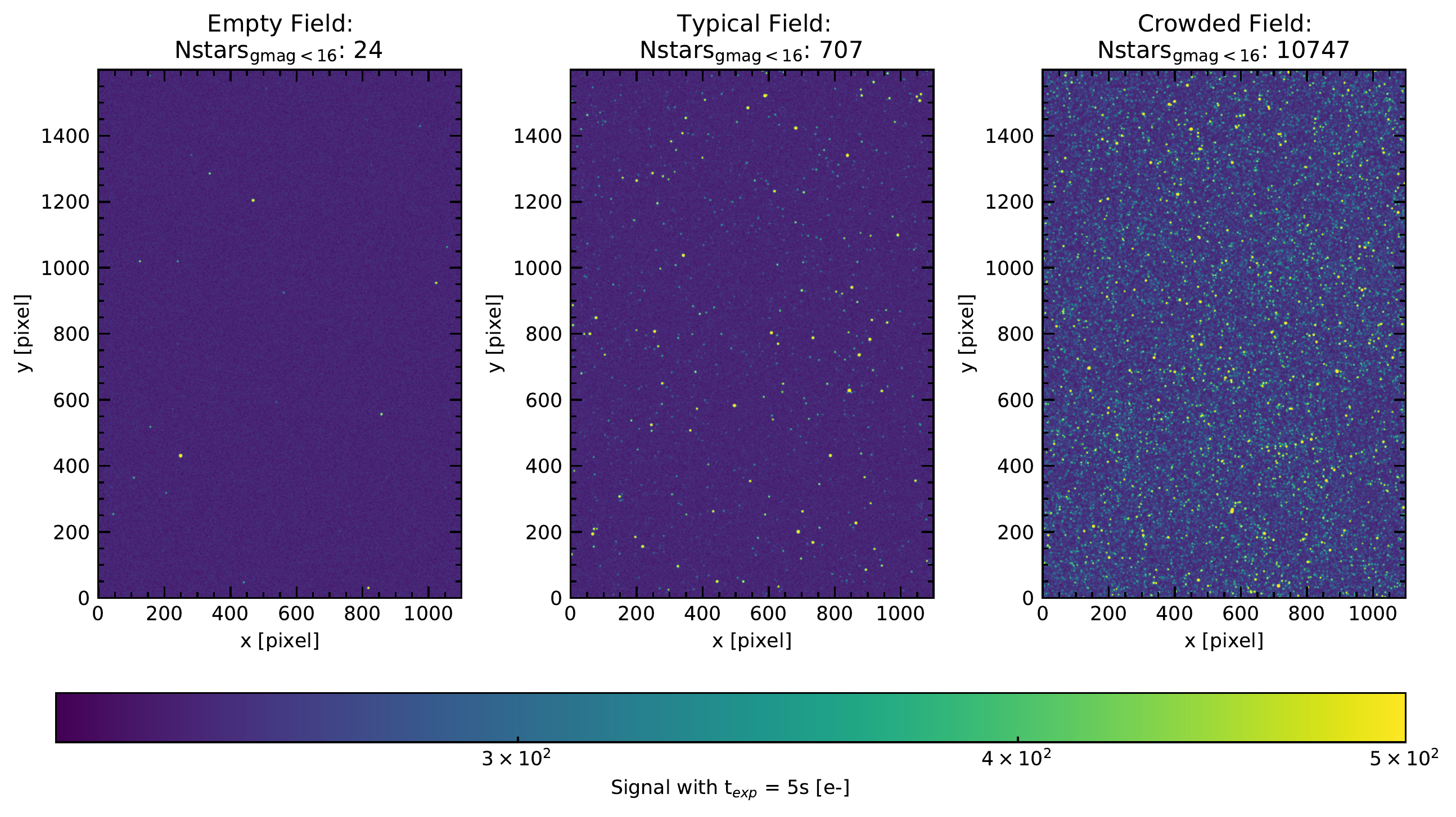}

   \end{center}
   \caption[synthetic image] 
   { \label{fig:synthetic_image} 
Three simulated guide camera images for (left to right) an empty, a typical, and a crowded field.}
   \end{figure}

To test and optimise the exposure time and the guiding software that will process the images taken with the guide cameras we simulated realistic exposures with the guide cameras. We assume Poisson shot noise for all photon/electron-counting quantities (stars, sky, dark current) and normal-distributed readout noise (with $\sigma_e=5.6e^-/s$, see \ref{subsec:noise}). We modelled the point-spread function as a Gaussian with a FWHM equal to our requirement (3.5~arcsec). The results for three different simulated images in fields with varying stellar density appear in Figure~\ref{fig:synthetic_image}.

\clearpage
\section{FOCAL PLANE METROLOGY}

\subsection{Motivation and requirements}
The focal plane metrology of the LVM telescopes has two phases and two goals:

In the first step, we will align both guide cameras and the IFU to a single reference plane. This will ensure that all these components have a common focus. The requirement for the coplanarity  50~µm is (derived from the rather slow speed of the f/11.42-beam and the requirements on the PSF width).

The goal of the second phase is to measure the exact relative positions. This is necessary to be able to place the IFU fibers on the correct sky positions, based on images taken with the A\&G cameras. The requirement of knowing the fibre position to 1~arcsec accuracy leads to a requirement of knowing the relative positions in the focal plane to less than 10~µm.
\subsection{Measurement setup}
We use a \textit{Point Source Microscope} (PSM)\cite{2005SPIE.5877..102P} from \textit{OPTICAL PERSPECTIVE GROUP } to measure the locations of both the microlenses of the IFU and the pixels of the CMOS sensors of the A\&G cameras. The PSM is an autostigmatic microscope that projects the output of a single mode fibre towards a point source in its focal plane. This point source can be used to probe the location of objects in the focal plane, as its location can be determined very accurately due to its small diameter (<5~µm) and its very short depth of focus. We use a long working-distance objective\footnote{Objective lens: \textit{10X Mitutoyo Plan Apo Infinity Corrected Long WD Objective}}, giving a working-distance of 34~mm. 

We mount the PSM on three precision linear stages equipped with \textit{Newport DMH-1} digital micrometer heads for fast and precise reading of the positions. The micrometers have a resolution of 1~µm and a stated accuracy of $\pm$2~µm over a range of 25.4~mm. The focal plane is wider than this (see Figure~\ref{fig:fp_layout}), so we extend the range of the micrometers using precisely manufactured extension blocks with a known width of either 16~mm or 32~mm. Figure~\ref{fig:psm_setup} shows pictures of the setup.

This setup allows us to measure 3D positions in the focal plane to high accuracy. Using the linear stages, we move the PSM over the edge of a microlens or a specific pixel of one of the camera sensors. By bringing the reflected image of the point source into focus (again with a linear stage), we can ensure that the PSM is always at the same distance from the component being measured.

To avoid having to count single pixels on the camera sensors, we read-out the guide cameras and locate the PSM spot. As the point-source is very small ($<$~5~µm) we can illuminate individual pixels.

As the the actual IFU (connected to the fibres) and the guiding components will be joined for the first time at the observatory, we used a bare microlens array with the same properties as the nominal microlenses for our tests. We built an adaptor to precisely place it at the right location in the focal plane. Other than that, all components are those that will be used in the final setup.
\subsection{Measurement procedure}
\begin{figure} [ht]
   \begin{center}
   
   \includegraphics[height=0.42\textwidth]{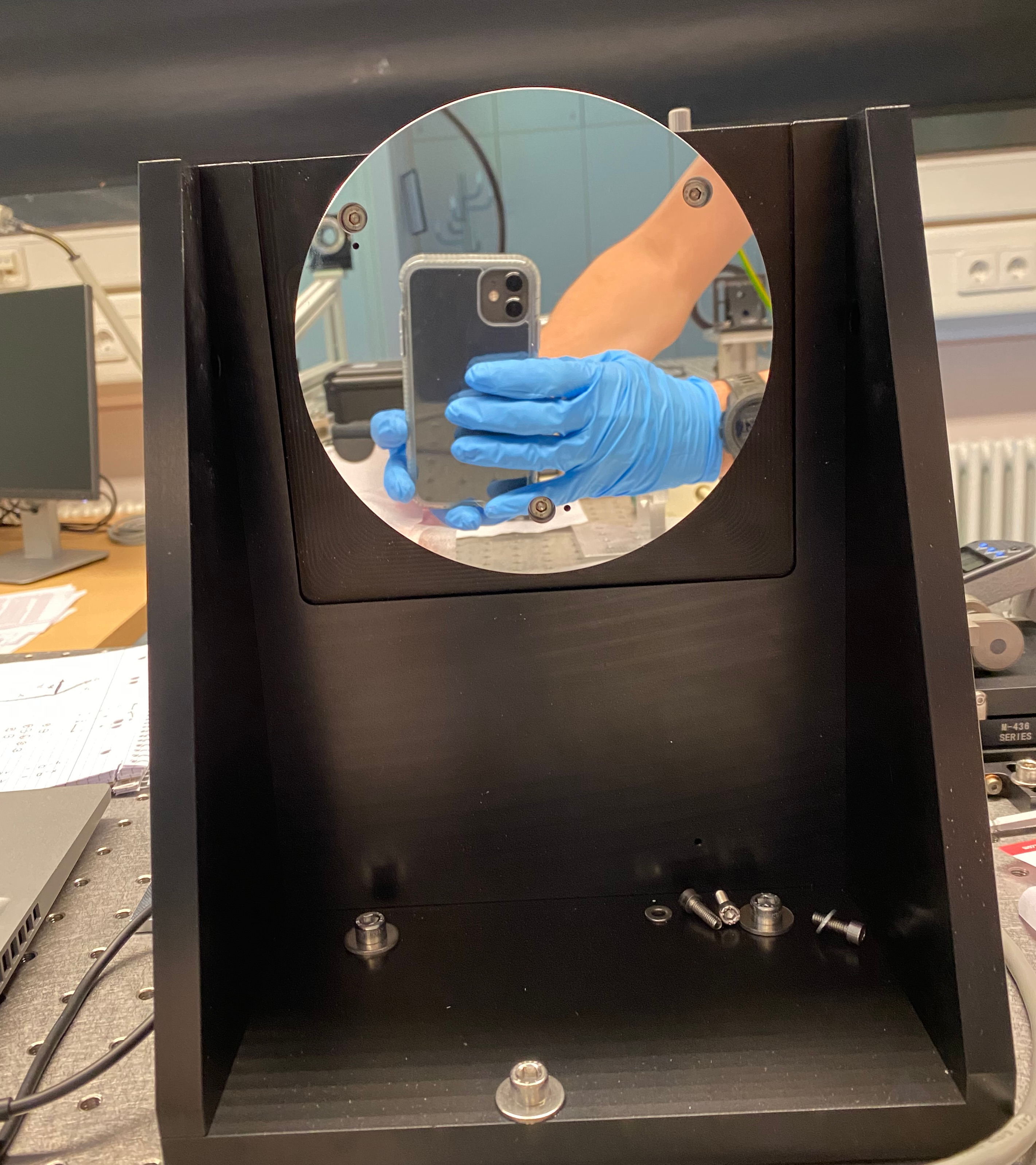}
   \includegraphics[height=0.42\textwidth]{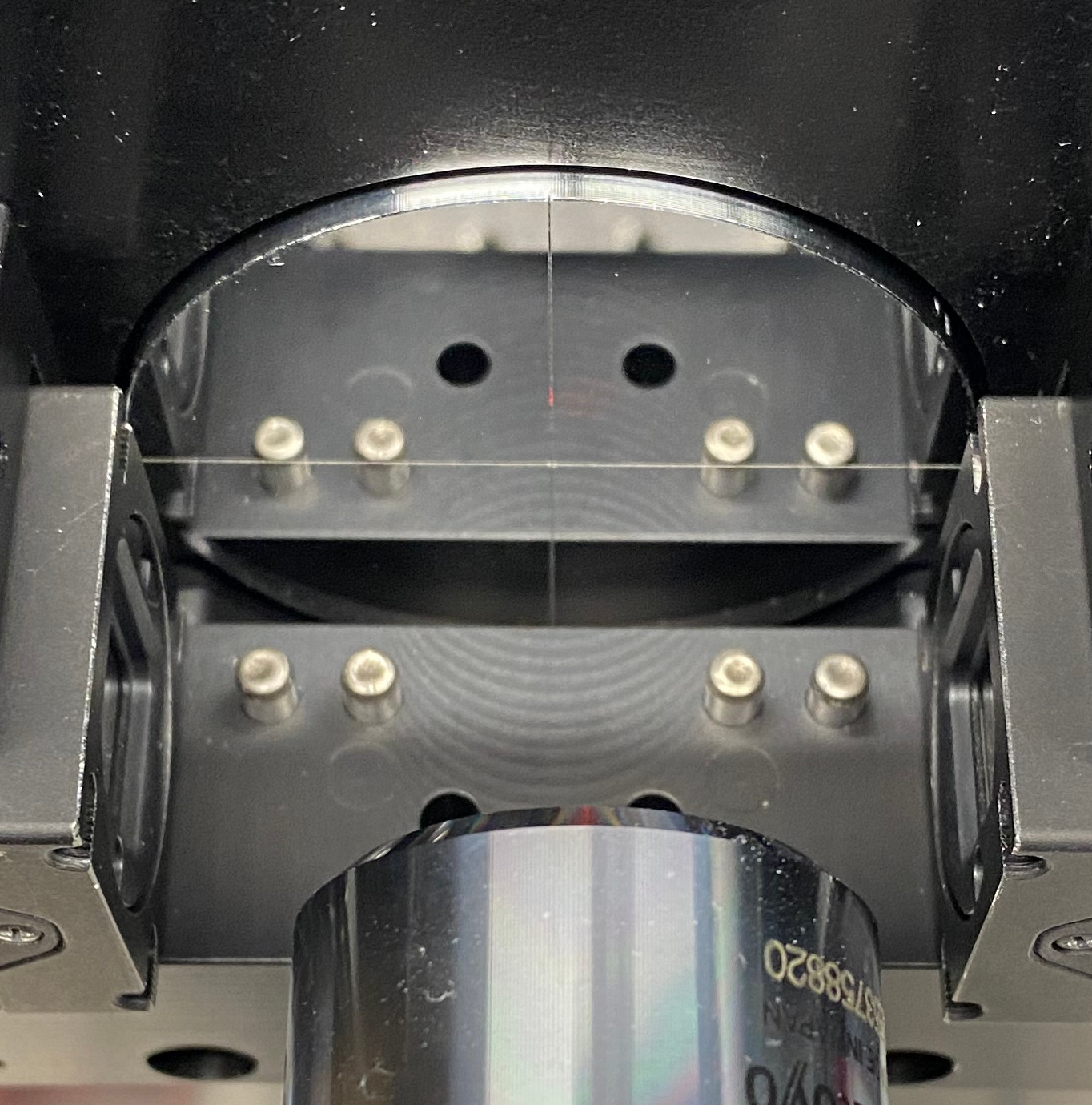}
   \includegraphics[width=0.8\textwidth]{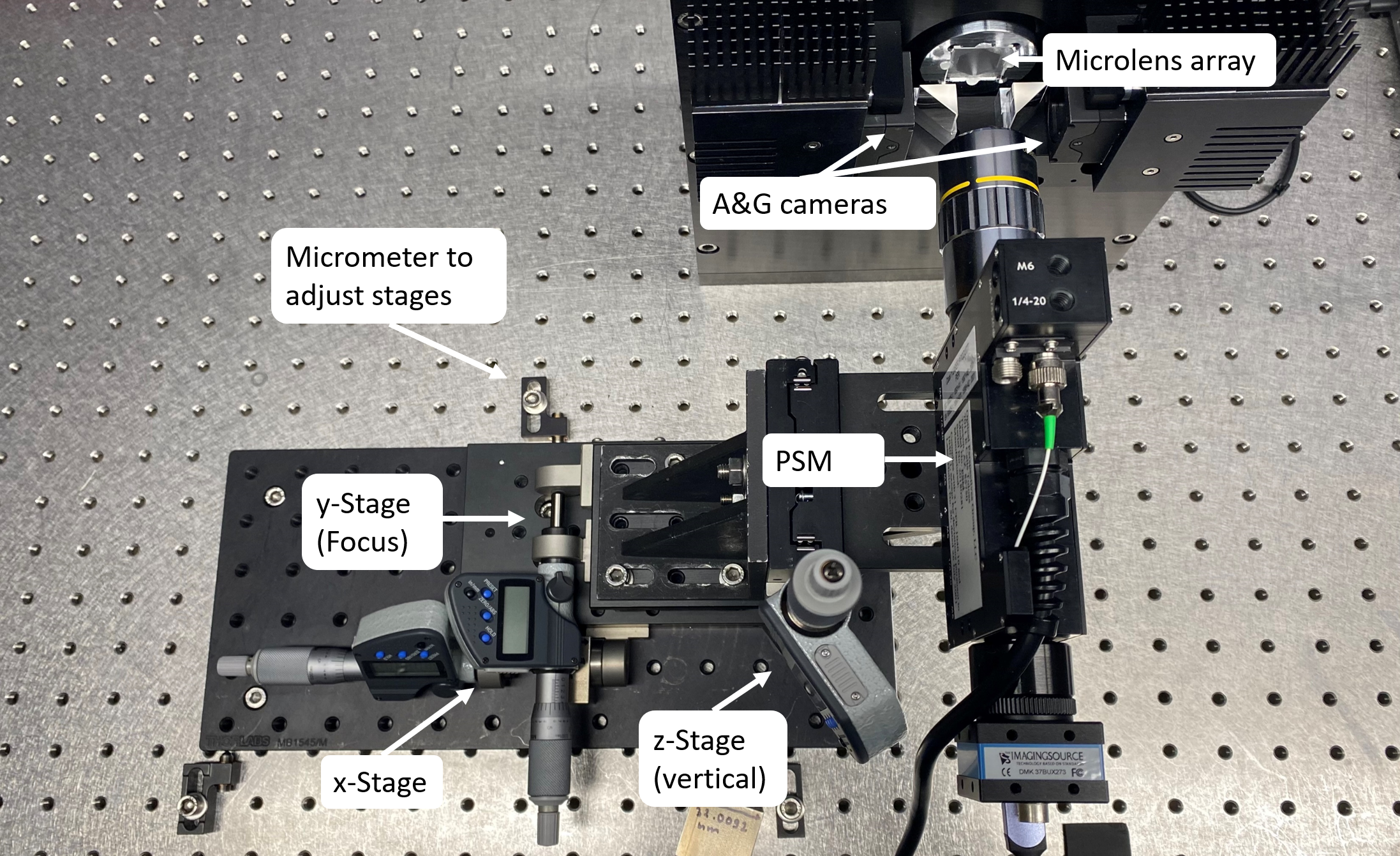}

   \end{center}
   \caption[psm setup] 
   { \label{fig:psm_setup} 
Top left: Backside of the focal plane alignment target. Top right: Front of the focal plane alignment target. One can also see the objective lens of the PSM. Bottom: Top-view of the focal plane metrology setup.}
   \end{figure}

The measurement procedure has multiple steps:
\begin{description}
    \item[1. Stage alignment] As a first step, we have to ensure that the stages are aligned perfectly with the desired focal plane. To do this, we install a diamond-turned target (see Figure~\ref{fig:psm_setup}), whose surface defines the focal plane. Then, we use micrometers mounted to the optical table to adjust the stage positions, before fixing the setup in place.
    \item[2. Hybrid measurements on camera sensors] After the stages have been fixed on the table, we remove the target from the focal plane and install the guide cameras. Then, we measure multiple positions on each of the 2 camera sensors. We identify which pixel is illuminated by the PSM by reading out the cameras. At the same time, we log the positions measured with the digital micrometers that move the stages. In principle, 3 measurements would be enough to fully determine the spatial position of each sensor, but to test the precision of our method and improve statistics, we take 9 measurements at different locations on each sensor.
    \item[3. Shimming of camera sensors] The guide cameras are also used to focus the telescope during operation. Therefore, both camera sensors and the microlens array of the IFU have to be in the same plane. If we determine a focus or tilt difference between the reference plane (set by the precisely turned focal plane target, see step 1) and the camera sensors, we will use shims to adjust the position of the cameras. This will likely be the case at the beginning due to manufacturing tolerances.
    \item[(Repetition of 2. and 3.)] These two steps are iterated, until the two sensors are in the correct plane. After that the as-built positions are measured on last time.
    \item[4. Measurement of as-built microlens positions] We need to locate the centre of each microlens. However, due to reflection and refraction, it is much easier to place the point source at the edge of a microlens. If we take three measurements at the edge of a particular microlens, we already can determine its centre (assuming it is perfectly circular), but again, to have a grasp of our measurement errors and better statistics, we take six measurements per microlens.
\end{description}

 \begin{figure} [ht]
   \begin{center}
   \includegraphics[height=5cm]{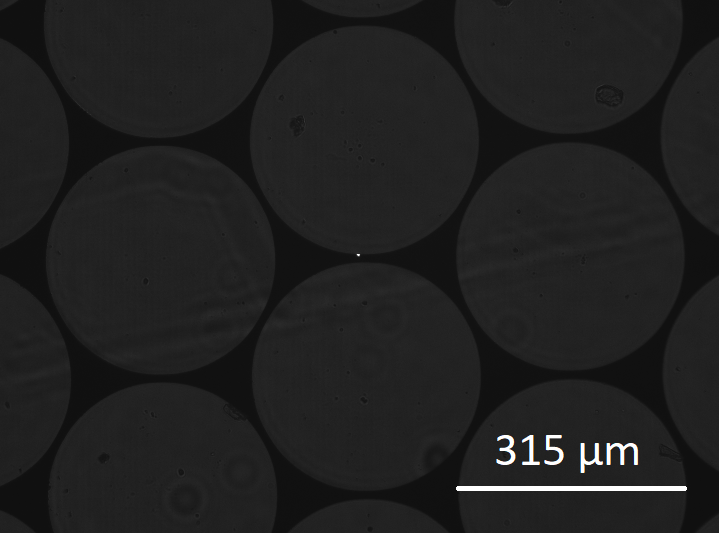}
   \includegraphics[height=5cm]{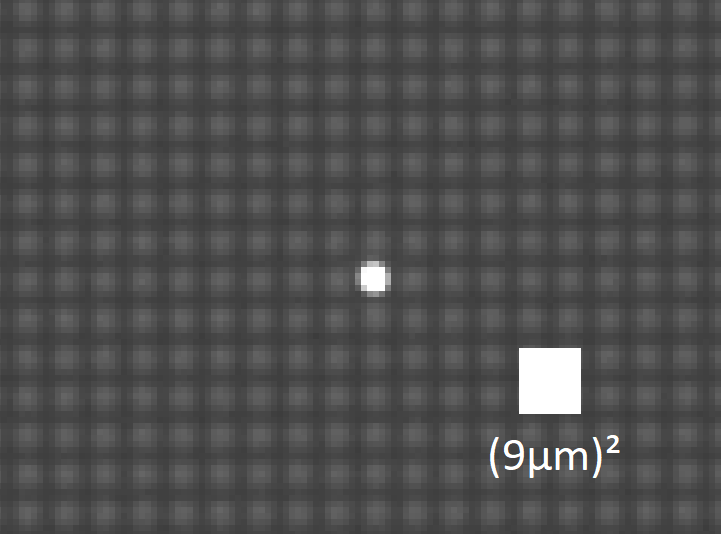}

   \end{center}
   \caption[psm setup] 
   { \label{fig:psm_view} 
The read out of the point source microscope. The left image shows the view of the microlenses, the right image shows (at higher digital magnification) the surface of one of the A\&G cameras. Due to the design of the sensor, each electronical pixel (size 9~µm) consists of 2$\times$2 physical units with a size of 4.5~µm. In both images, the point source can be seen as a bright spot in the center.}
   \end{figure}

\subsection{Data reduction}
While the three stages directly measure physical positions in the focal plane, we still have to use some mathematics to fully understand the focal plane geometry.

The positions of the camera are determined by finding the 7-parameter linear transformations (3 translations, 3 rotations, scale), that transform pixel coordinates into physical 3D coordinates using a least-squares fit. 

The positions of the microlenses are determined by fitting a circle to each of the 6 positions measurements of the edge of each measured lens. After the positions of a few of the individual lenses have been determined, we can  determine the global position and orientation of the full microlens array with another linear transformation.

\subsection{Results}
While tests are still ongoing, the first results of this metrology setup are very promising. By comparing the measured physical positions with the on-chip astrometry (i.e. the identified A\&G pixels), we can test the consistency of the measurements. The residuals show a RMS of 2~µm (with a P-V of 10~µm), which is consistent with the accuracy of the micrometer heads. This also demonstrates that we are achieving the required precision level.

\begin{figure} [ht]
   \begin{center}
   \includegraphics[width=1.0\textwidth]{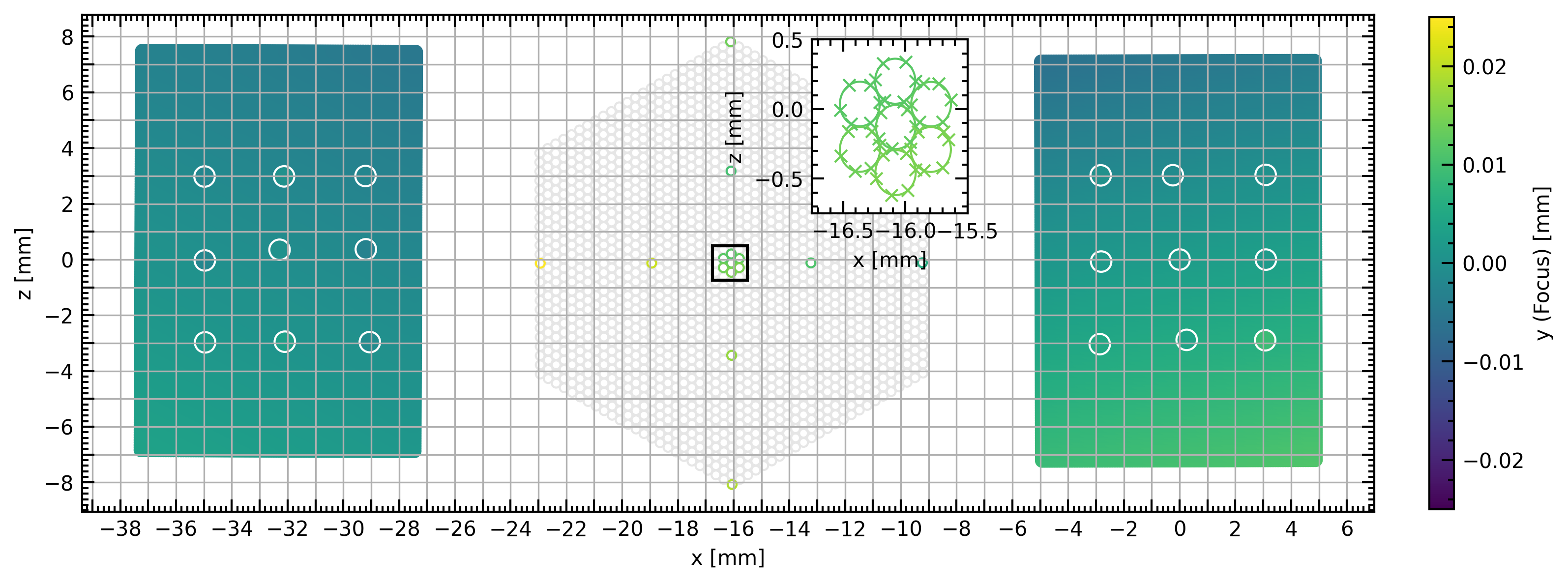}

   \end{center}
   \caption[psm results] 
   { \label{fig:psm_results} 
The as-built focal plane with positions determined using our high precision metrology setup.  The two large rectangles show the inferred positions of the CMOS sensors, colour-coded with the focus deviation, while the hexagon in the middle shows the location of the IFU. Colored circles indicate microlenses, that have been measured. The small insert shows a magnification of the central region of the IFU, the individual measurements taken at the edge of the lenslets are marked with crosses.}
   \end{figure}

The coordinates and transformations determined with the PSM measurements were used to accurately calculate the required adjustments to the shims used to align the A\&G cameras to the reference focal plane. This proved to be very effective - in 2 iterations of re-shimming we reduced the RMS of the focus deviations from initial $\sim$0.6-0.7~mm to to $\leq$0.005~mm (see Table~\ref{tab:shimming_results}), clearly matching our focus requirement of 0.05~mm. Figure~\ref{fig:psm_results} shows the as-built geometry of our test focal plane as measured with the PSM setup after the 2 iterations of shimming, while Table~\ref{tab:metrology_results} lists the numerical results. 

\begin{table}[ht]
\caption{Focus deviations for two cameras during different iterations of the shimming process.} 
\label{tab:shimming_results}
\begin{center}       
\begin{tabular}{l|llll|llll|}
\cline{2-9}
                                                      & \multicolumn{4}{l|}{Camera 1}                 & \multicolumn{4}{l|}{Camera 2}                 \\ \cline{2-9} 
                                                      & \multicolumn{4}{l|}{Focus Deviation {[}mm{]}} & \multicolumn{4}{l|}{Focus Deviation {[}mm{]}} \\
                                                      & Min.       & Mean.     & Max.      & RMS      & Min.       & Mean.     & Max.      & RMS      \\ \hline
\multicolumn{1}{|l|}{Iteration 0 (standard shims)}    & 0.598      & 0.726     & 0.855     & 0.729    & 0.415      & 0.622     & 0.830     & 0.630    \\
\multicolumn{1}{|l|}{Iteration 1 (custom made shims)} & -0.060     & -0.046    & -0.032    & 0.046    & -0.136     & -0.086    & -0.036    & 0.089    \\
\multicolumn{1}{|l|}{Iteration 2 (custom made shims)} & -0.005     & -0.001    & 0.004     & 0.002    & -0.006     & 0.003     & 0.011     & 0.005    \\ \hline
\end{tabular}
\end{center}
\end{table}

\begin{table}[ht]
\caption{Preliminary results of the focal plane metrology, after the cameras have been shimmed to adjust for focus deviations.}
\label{tab:metrology_results}
\begin{center}       
\begin{tabular}{|p{2.8cm}|p{1.8cm}|p{1.6cm}|p{1.6cm}|p{1.4cm}|p{1.4cm}|p{1.4cm}|p{1.8cm}|}
\hline
\rule[-1ex]{0pt}{3.5ex} \textbf{Component}& \textbf{x-centre} & \textbf{y-centre (focus)} & \textbf{z-centre} & \textbf{Rotation x~axis} & \textbf{Rotation y~axis} & \textbf{Rotation z~axis} & \textbf{Scale} \\
\hline
\rule[-1ex]{0pt}{3.5ex}  Guide camera 1 & -32.286~mm & -0.001~mm &0.351~mm & -0.013$^\circ$ & 0.026$^\circ$ & -0.209$^\circ$ & 9~µm/pixel  \\
\rule[-1ex]{0pt}{3.5ex}  Guide camera 2 & 0.002~mm & 0.003~mm &-0.001~mm & 0.013$^\circ$ & 0.060$^\circ$ & 0.136$^\circ$ & 9~µm/pixel  \\
\rule[-1ex]{0pt}{3.5ex}  Microlens array & -16.079~mm & 0.015~mm &-0.128~mm & -0.089$^\circ$ & 0.025$^\circ$ & 0.141$^\circ$ & 330~µm/lens  \\

\hline
\end{tabular}
\end{center}
\end{table}
 
\section{SUMMARY}
In this work, we demonstrated the required performance of the LVM acquisition and guiding hardware.

We show that relatively low-cost commercial CMOS cameras offer sky-brightness limited performance, and we reach a sufficient depth for the LVM survey A\&G strategy. We reach a magnitude limit of $\sim$16.5 in 5~s in a full moon night. Even with a magnitude limit of 15 and only a single A\&G camera, there are always more than 5 stars available for guiding.

In addition, we present a metrology setup using a point source microscope, which allows us to measure the relative positions of the microlenses of the IFU and the pixels of the guide camera sensors to an accuracy of 2~µm in all three spatial dimensions, allowing for precise focus and tip/tilt adjustments.
\clearpage
\acknowledgments 
Funding for the Sloan Digital Sky Survey V has been provided by the Alfred P. Sloan Foundation, the Heising-Simons Foundation, and the Participating Institutions. SDSS acknowledges support and resources from the Center for High-Performance Computing at the University of Utah. The SDSS web site is \url{www.sdss5.org}.

SDSS is managed by the Astrophysical Research Consortium for the Participating Institutions of the SDSS Collaboration, including the Carnegie Institution for Science, Chilean National Time Allocation Committee (CNTAC) ratified researchers, the Gotham Participation Group, Harvard University, Heidelberg University, The Johns Hopkins University, L'Ecole polytechnique f\'{e}d\'{e}rale de Lausanne (EPFL), Leibniz-Institut f\"{u}r Astrophysik Potsdam (AIP), Max-Planck-Institut f\"{u}r Astronomie (MPIA Heidelberg), Max-Planck-Institut f\"{u}r Extraterrestrische Physik (MPE), Nanjing University, National Astronomical Observatories of China (NAOC), New Mexico State University, The Ohio State University, Pennsylvania State University, Smithsonian Astrophysical Observatory, Space Telescope Science Institute (STScI), the Stellar Astrophysics Participation Group, Universidad Nacional Aut\'{o}noma de M\'{e}xico, University of Arizona, University of Colorado Boulder, University of Illinois at Urbana-Champaign, University of Toronto, University of Utah, University of Virginia, Yale University, and Yunnan University.

This work has made use of data from the European Space Agency (ESA) mission
{\it Gaia} (\url{https://www.cosmos.esa.int/gaia}), processed by the {\it Gaia}
Data Processing and Analysis Consortium (DPAC,
\url{https://www.cosmos.esa.int/web/gaia/dpac/consortium}). Funding for the DPAC
has been provided by national institutions, in particular the institutions
participating in the {\it Gaia} Multilateral Agreement.

Some of the results in this paper have been derived using the healpy and HEALPix package
\bibliography{report} 

\begin{thebibliography}{10}

\bibitem{2017arXiv171103234K}
{Kollmeier}, J.~A., {Zasowski}, G., {Rix}, H.-W., {Johns}, M., {Anderson},
  S.~F., {Drory}, N., {Johnson}, J.~A., {Pogge}, R.~W., {Bird}, J.~C., {Blanc},
  G.~A., {Brownstein}, J.~R., {Crane}, J.~D., {De Lee}, N.~M., {Klaene}, M.~A.,
  {Kreckel}, K., {MacDonald}, N., {Merloni}, A., {Ness}, M.~K., {O'Brien}, T.,
  {Sanchez-Gallego}, J.~R., {Sayres}, C.~C., {Shen}, Y., {Thakar}, A.~R.,
  {Tkachenko}, A., {Aerts}, C., {Blanton}, M.~R., {Eisenstein}, D.~J.,
  {Holtzman}, J.~A., {Maoz}, D., {Nandra}, K., {Rockosi}, C., {Weinberg},
  D.~H., {Bovy}, J., {Casey}, A.~R., {Chaname}, J., {Clerc}, N., {Conroy}, C.,
  {Eracleous}, M., {G{\"a}nsicke}, B.~T., {Hekker}, S., {Horne}, K.,
  {Kauffmann}, J., {McQuinn}, K. B.~W., {Pellegrini}, E.~W., {Schinnerer}, E.,
  {Schlafly}, E.~F., {Schwope}, A.~D., {Seibert}, M., {Teske}, J.~K., and {van
  Saders}, J.~L., ``{SDSS-V: Pioneering Panoptic Spectroscopy},'' {\em arXiv
  e-prints} ,  arXiv:1711.03234 (Nov. 2017).

\bibitem{2020SPIE11447E..18K}
{Konidaris}, N.~P., {Drory}, N., {Froning}, C.~S., {Hebert}, A., {Bilgi}, P.,
  {Blanc}, G.~A., {Lanz}, A.~E., {Hull}, C.~L., {Kollmeier}, J.~A., {Ramirez},
  S., {Wachter}, S., {Kreckel}, K., {Pak}, S., {Pellegrini}, E., {Almeida}, A.,
  {Case}, S., {Zhelem}, R., {Feger}, T., {Lawrence}, J., {Lesser}, M.,
  {Herbst}, T., {Sanchez-Gallego}, J., {Bershady}, M.~A., {Chattopadhyay}, S.,
  {Hauser}, A., {Smith}, M., {Wolf}, M.~J., and {Yan}, R., ``{SDSS-V local
  volume mapper instrument: overview and status},'' in [{\em Society of
  Photo-Optical Instrumentation Engineers (SPIE) Conference
  Series}{\nolinebreak\hspace{0.1em}]},  {\em Society of Photo-Optical
  Instrumentation Engineers (SPIE) Conference Series} {\bf 11447},  1144718
  (Dec. 2020).

\bibitem{2020SPIE11445E..0JH}
{Herbst}, T.~M., {Bilgi}, P., {Bizenberger}, P., {Blanc}, G., {Briegel}, F.,
  {Case}, S., {Drory}, N., {Feger}, T., {Froning}, C., {Gaessler}, W.,
  {Hebert}, A., {Konidaris}, N., {Lanz}, A., {Mohr}, L., {Pak}, S.,
  {Ram{\'\i}rez}, S., {Rohloff}, R.-R., {S{\'a}nchez-Gallego}, J., and
  {Wachter}, S., ``{The SDSS-V local volume mapper telescope system},'' in
  [{\em Society of Photo-Optical Instrumentation Engineers (SPIE) Conference
  Series}{\nolinebreak\hspace{0.1em}]},  {\em Society of Photo-Optical
  Instrumentation Engineers (SPIE) Conference Series} {\bf 11445},  114450J
  (Dec. 2020).

\bibitem{2018SPIE10702E..7KP}
{Perruchot}, S., {Guy}, J., {Le Guillou}, L., {Blanc}, P.~E., {Ronayette}, S.,
  {R{\'e}gal}, X., {Castagnoli}, G., {Sepulveda}, E., {Le Van Suu}, A.,
  {Jullo}, E., {Cuby}, J.~G., {Karkar}, S., {Ghislain}, P., {Repain}, P.,
  {Carton}, P.~H., {Magneville}, C., {Ealet}, A., {Escoffier}, S., {Secroun},
  A., {Cousinou}, M.~C., {Honscheid}, K., {Elliot}, A., {Jelinsky}, P.,
  {Brooks}, D., and {Tarl{\`e}}, G., ``{Integration and testing of the DESI
  multi-object spectrograph: performance tests and results for the first unit
  out of ten},'' in [{\em Ground-based and Airborne Instrumentation for
  Astronomy VII}{\nolinebreak\hspace{0.1em}]},  {Evans}, C.~J., {Simard}, L.,
  and {Takami}, H., eds., {\em Society of Photo-Optical Instrumentation
  Engineers (SPIE) Conference Series} {\bf 10702},  107027K (July 2018).

\bibitem{2006SPIE.6276E..0MJ}
{Janesick}, J., {Andrews}, J.~T., and {Elliott}, T., ``{Fundamental performance
  differences between CMOS and CCD imagers: Part 1},'' in [{\em Society of
  Photo-Optical Instrumentation Engineers (SPIE) Conference
  Series}{\nolinebreak\hspace{0.1em}]},  {Dorn}, D.~A. and {Holland}, A.~D.,
  eds., {\em Society of Photo-Optical Instrumentation Engineers (SPIE)
  Conference Series} {\bf 6276},  62760M (June 2006).

\bibitem{2013SPIE.8659E..02J}
{Janesick}, J.~R., {Elliott}, T., {Andrews}, J., {Tower}, J., and {Pinter}, J.,
  ``{Fundamental performance differences of CMOS and CCD imagers: part V},'' in
  [{\em Sensors, Cameras, and Systems for Industrial and Scientific
  Applications XIV}{\nolinebreak\hspace{0.1em}]},  {Widenhorn}, R. and
  {Dupret}, A., eds., {\em Society of Photo-Optical Instrumentation Engineers
  (SPIE) Conference Series} {\bf 8659},  865902 (Feb. 2013).

\bibitem{2017JInst..12C7008J}
{Jorden}, P.~R., {Jerram}, P.~A., {Fryer}, M., and {Stefanov}, K.~D., ``{e2v
  CMOS and CCD sensors and systems for astronomy},'' {\em Journal of
  Instrumentation}~{\bf 12},  C07008 (July 2017).

\bibitem{2021A&A...649A...1G}
{Gaia Collaboration}, {Brown}, A.~G.~A., and {Vallenari}, A., ``{Gaia Early
  Data Release 3. Summary of the contents and survey properties},'' {\em
  A\&A}~{\bf 649},  A1 (May 2021).

\bibitem{2021A&A...649A...5F}
{Fabricius}, C., {Luri}, X., {Arenou}, F., {Babusiaux}, C., {Helmi}, A.,
  {Muraveva}, T., {Reyl{\'e}}, C., {Spoto}, F., {Vallenari}, A., {Antoja}, T.,
  {Balbinot}, E., {Barache}, C., {Bauchet}, N., {Bragaglia}, A., {Busonero},
  D., {Cantat-Gaudin}, T., {Carrasco}, J.~M., {Diakit{\'e}}, S., {Fabrizio},
  M., {Figueras}, F., {Garcia-Gutierrez}, A., {Garofalo}, A., {Jordi}, C.,
  {Kervella}, P., {Khanna}, S., {Leclerc}, N., {Licata}, E., {Lambert}, S.,
  {Marrese}, P.~M., {Masip}, A., {Ramos}, P., {Robichon}, N., {Robin}, A.~C.,
  {Romero-G{\'o}mez}, M., {Rubele}, S., and {Weiler}, M., ``{Gaia Early Data
  Release 3. Catalogue validation},'' {\em A\&A}~{\bf 649},  A5 (May 2021).

\bibitem{Zonca2019}
Zonca, A., Singer, L., Lenz, D., Reinecke, M., Rosset, C., Hivon, E., and
  Gorski, K., ``healpy: equal area pixelization and spherical harmonics
  transforms for data on the sphere in python,'' {\em Journal of Open Source
  Software}~{\bf 4},  1298 (Mar. 2019).

\bibitem{2005ApJ...622..759G}
{G{\'o}rski}, K.~M., {Hivon}, E., {Banday}, A.~J., {Wandelt}, B.~D., {Hansen},
  F.~K., {Reinecke}, M., and {Bartelmann}, M., ``{HEALPix: A Framework for
  High-Resolution Discretization and Fast Analysis of Data Distributed on the
  Sphere},'' {\em The Astrophysical Journal}~{\bf 622},  759--771 (Apr. 2005).

\bibitem{2005SPIE.5877..102P}
{Parks}, R.~E. and {Kuhn}, W.~P., ``{Optical alignment using the Point Source
  Microscope},'' in [{\em Optomechanics 2005}{\nolinebreak\hspace{0.1em}]},
  {Hatheway}, A.~E., ed., {\em Society of Photo-Optical Instrumentation
  Engineers (SPIE) Conference Series} {\bf 5877},  102--116 (Aug. 2005).

\end{thebibliography}
\bibliographystyle{spiebib} 

\end{document}